\documentclass[10pt]{article}

\newlength{\minitwocolumn}
\font\teneufm=eufm10
\font\seveneufm=eufm7
\font\fiveeufm=eufm5
\newfam\eufmfam
\textfont\eufmfam=\teneufm
\scriptfont\eufmfam=\seveneufm
\scriptscriptfont\eufmfam=\fiveeufm

\makeatletter
\@addtoreset{equation}{section}
\makeatother

\title{\bf
\Large{\bf 
Correlation functions of the half-infinite
XXZ spin chain \\
with a triangular boundary}}
\begin{document}

\maketitle

\begin{center}
{P. Baseilhac$~^{\alpha}$} and {T. Kojima$~^{\beta}$}
\\~\\
{\it {$\alpha$}~
Laboratoire de Math\'ematiques et Physique Th\'eorique CNRS/UMR 7350,
\\F\'ed\'eration Denis Poisson FR2964,
Universit\'e de Tours,
Parc de Grammont, 37200 Tours, 
FRANCE\\
baseilha@lmpt.univ-tours.fr
\\
~
\\
{$\beta$}~
Department of Mathematics and Physics,
Faculty of Engineering,
Yamagata University,\\
 Jonan 4-3-16, Yonezawa 992-8510, JAPAN\\
kojima@yz.yamagata-u.ac.jp}
\end{center}

~\\
\\

\begin{abstract}
The half-infinite XXZ spin chain with a triangular boundary
is considered in the massive regime. 
Two integral representations of 
correlation functions are proposed using bosonization. 
Sufficient conditions such that the expressions for triangular boundary conditions 
coincide with 
those for diagonal boundary conditions
are identified. 
As an application, 
summation formulae of
the boundary expectation values 
$\langle \sigma_1^{a}\rangle $ with $a=z,\pm$ are obtained.
Exploiting the spin-reversal property,
relations between $n$-fold integrals of elliptic theta functions are extracted.
\end{abstract}

~\\

\newpage

\section{Introduction}

Beyond the Ising model,
a large class of solvable lattice models have been discovered.
In the context of quantum integrable systems on the lattice,
spin chains are among the most studied examples with applications which range from condensed matter to
high energy physics.
Given a Hamiltonian, finding analytical expressions for the exact spectrum, 
identifying the structure of the space of the eigenvectors
and deriving explicit expressions for correlation functions 
are essential steps in the non-perturbative  characterization of the system's behavior which can be
compared with experimental data. 

Among the simplest examples considered in the literature, the $XXZ$ spin chain
with different boundary conditions has received particular attention.
Over the years, different approaches have been proposed in order to understand the Hamiltonian's spectral problem
and derive the correlation functions.
For models with periodic boundary conditions,
the spectral problem can be handled by methods such as
the Bethe ansatz (BA) \cite{Bethe},
or the corner transfer matrix method (CTM) in the thermodynamic limit \cite{Baxter}.
The computation of the correlation functions, however, is a much more difficult problem in general.
Apart from the simplest example - namely the $XXZ$ spin chain with periodic boundary condition -
for which correlation functions have been proposed by the quantum inverse scattering method (QISM)
\cite{KBI, KMT} arising from the BA,
the generalization of this result to models with higher symmetries requires a better understanding of 
mathematical structures, for instance,  of determinant formulae of scalar products that involve the Bethe vectors 
\cite{Reshetikhin} (see some recent progress in \cite{Wheeler}).
However, in the thermodynamic limit, this problem can be alternatively tackled using the $q$-vertex operator approach (VOA) \cite{DFJMN}
arising from the CTM.
The space of states is identified with the irreducible highest weight representation of 
$U_q(\widehat{sl_2})$ or higher rank quantum algebras.
Correlation functions can be obtained using bosonizations of the $q$-vertex operators for $U_q(\widehat{sl_2})$ 
or higher rank quantum algebras
\cite{DFJMN, FJ, Koyama, JKK, Bernard, KSU}.
Either within the QISM or the VOA, correlation functions 
are obtained in the form of integrals of meromorphic functions
in the thermodynamic limit.

The situation for integrable spin chains with open boundaries is more difficult. 
On one hand, for the finite $XXZ$ open chain with diagonal boundaries \cite{Sklyanin},
related non-diagonal boundaries \cite{Nepomechie,CLSW}
or $q$ a root of unity \cite{MNS}, 
the BA makes it possible to derive the spectrum and the eigenvectors \footnote{Note that recently, a modified BA approach has been considered which looks promising \cite{Cao2,BC}.} \cite{CLSW}. In each case, the corresponding models are studied using Sklyanin's general formulation of the BA applied to open boundary models \cite{Sklyanin}.
On the other hand, in the thermodynamic limit, the VOA has been applied to 
the half-infinite $XXZ$ spin chain with a diagonal boundary \cite{JKKKM}.
Although the hidden symmetry of this model was still unknown at that time\footnote{Recently, the hidden symmetry of the spin chain with a diagonal and non-diagonal boundary has been identified: it is associated with the augmented $q$-Onsager algebra and $q$-Onsager algebra, respectively \cite{BB1}.}, the diagonalization of the Hamiltonian could still be achieved. Based on these results, the computation of
correlation functions has been achieved for diagonal boundary conditions either using the BA \cite{KZMNST} or using the VOA in the thermodynamic limit \cite{JKKKM}. Note that generalizations to models with higher symmetries have been studied for 
$U_q(\widehat{sl_N})$ etc. \cite{FK, Kojima1, Kojima2}. In the thermodynamic limit, when the comparison is feasible the expressions obtained by both approaches essentially coincide.

In spite of these important developments, the computation of correlation functions of the $XXZ$ open spin chain for more general boundary conditions has remained, up to now, essentially problematic. On one hand, even in a simpler case such as the $XXZ$ spin chain with triangular boundary conditions for which the construction of the Bethe vector is feasible \cite{BCR,PL}, it remains an open problem in the QISM.  On the other hand,  in the thermodynamic limit the application of the VOA requires the prior knowledge of the vacuum eigenvectors of the Hamiltonian. Since 1994 \cite{JKKKM}, even for the simpler case of triangular boundary conditions the solution to this problem has been unknown. However, a breakthrough was recently made in \cite{BB1}, which has opened the possibility of computing correlation functions in the thermodynamic limit of the XXZ open spin chain. Namely, based on the so-called Onsager's approach the structure of the eigenvectors of the finite $XXZ$ spin chain for any type of boundary conditions was interpreted within the representation theory of the $q$-Onsager algebra \cite{BK,BB1}.
In the thermodynamic limit, the $q$-Onsager algebra is realized by quadratics of the $q$-vertex operators associated with $U_q(\widehat{sl_2})$ \cite{BB1} (see also \cite{BB2} for an alternative derivation). As a consequence, vacuum eigenvectors of the Hamiltonian for a triangular boundary were constructed using the intertwining properties of the $q$-vertex operators with monomials of the $q$-Onsager basic generators \cite{BB1}.  The latter being expressed in terms of $U_q(\widehat{sl_2})$ generators, the VOA can be applied in a straightforward manner. 

The purpose of this paper is to present the first examples of correlation functions of the half-infinite  XXZ open spin chain with a {\it non-diagonal} boundary, using the framework of the VOA. The results here presented extend the earlier studies \cite{JKKKM,BB1}. Among the  applications, closed formulae for the boundary expectation values of  the spin operators are given and remarkable identities between $n-$fold integrals of elliptic theta functions are exhibited. Here we focus our attention on the simplest non-diagonal example, namely the half-infinite $XXZ$ spin chain with upper or lower triangular boundary condition.  We are interested in the Hamiltonian :
\begin{eqnarray}
H_B^{(\pm)}=-\frac{1}{2}
\sum_{k=1}^\infty (\sigma_{k+1}^x \sigma_k^x+\sigma_{k+1}^y \sigma_k^y+\Delta \sigma_{k+1}^z \sigma_k^z)
-\frac{1-q^2}{4q}\frac{1+r}{1-r}\sigma_1^z-\frac{s}{1-r} \sigma_1^\pm,
\label{def:Hamiltonian}
\end{eqnarray}
where we have used the standard Pauli matrices
\begin{eqnarray}
\sigma^x=\left(\begin{array}{cc}
0&1\\
1&0
\end{array}\right),~
\sigma^y=
\left(\begin{array}{cc}
0&-i\\
i&0
\end{array}\right),~
\sigma^z=
\left(\begin{array}{cc}
1&0\\
0&-1
\end{array}\right),~
\sigma^+=\left(\begin{array}{cc}
0&1\\
0&0
\end{array}\right),~
\sigma^-=\left(\begin{array}{cc}
0&0\\
1&0
\end{array}\right).
\end{eqnarray}
Here we consider the model in the limit of the half-infinite spin chain, in the massive regime where
\begin{eqnarray}
\Delta=\frac{q+q^{-1}}{2},~~~-1<q<0,~~~-1\leq r \leq 1,~~~s \in {\bf R}.
\end{eqnarray}
Since under conjugation of $H_B^{(\pm)}$ by the spin-reversal operator 
$\hat{\nu}=\prod_{j=1}^\infty \sigma_j^x$
the sign of the boundary term is reversed, we can restrict our discussion to the boundary term
$-\frac{1-q^2}{4q}\frac{1+r}{1-r}
\geq 0$, or $-1 \leq r \leq 1$. Importantly,
the two fundamental vacuum eigenvectors $|\pm ; i \rangle_B~(i=0,1)$ for 
the triangular boundary models $H_B^{(\pm)}$ 
were constructed in a recent paper \cite{BB1}.
For instance, for the lower triangular boundary model $H_B^{(-)}$, 
the two fundamental vacuum eigenvectors\footnote{By definition \cite{BB1}, $|\pm ; i \rangle_B~(i=0,1)$ are eigenvectors of the Hamiltonian (\ref{def:Hamiltonian}), not to be confused with the objects called the {\it pseudo-vacuum} vectors that arise in the algebraic BA approach.} are given by $q$-exponentials of $U_q(\widehat{sl_2})$ Chevalley generators acting on the vacuum eigenvectors  $|i\rangle_B~(i=0,1)$  of the model with a diagonal boundary $(s=0)$ \cite{JKKKM}:
\begin{eqnarray}
|-;0\rangle_B=\exp_{q}\left(-\frac{~s~}{q}e_0 q^{-h_0}\right)|0\rangle_B,
~~~
|-;1\rangle_B=\exp_{q^{-1}}\left(-\frac{~s~}{r}f_1\right)|1\rangle_B.
\end{eqnarray}
Here we have used the $q$-exponential function
\begin{eqnarray}
\exp_q(x)=\sum_{n=0}^\infty \frac{q^{\frac{n(n-1)}{2}}}{[n]_q!}x^n.
\label{def:q-exp}
\end{eqnarray}
In this paper we give the dual vacuum eigenvectors $~_B\langle i;\pm|~(i=0,1)$ using the intertwining properties of 
the $q$-vertex operators of $U_q(\widehat{sl_2})$.
For instance, for the  lower triangular boundary model $H_B^{(-)}$, 
the two fundamental dual vacuum eigenvectors are given by
\begin{eqnarray}
~_B\langle 0;-|=~_B\langle 0|\exp_{q^{-1}}\left(\frac{~s~}{q}e_0q^{-h_0}\right),
~~~_B\langle 1;-|=~_B\langle 1|
\exp_{q}\left(\frac{~s~}{r}f_1\right).
\end{eqnarray}
Using these, we compute the integral representations of the correlation functions using the bosonizations.
As a special case, the summation formulae of the boundary expectation values of the spin operators are derived:
\begin{eqnarray}
\frac{~_B\langle 0;-|\sigma_1^z |-;0\rangle_B}{~_B\langle 0;-|-;0\rangle_B}
&=&-1-2(1-r)^2 \sum_{n=1}^\infty
\frac{(-q^2)^n}{(1-rq^{2n})^2},\\
\frac{~_B\langle 0;-|\sigma_1^+ |-;0\rangle_B}{~_B\langle 0;-|-;0\rangle_B}
&=&s
\left(2+(1-r)\sum_{n=1}^\infty
(-q^2)^n\frac{2q^{2n}-r(1+q^{4n})}{(1-rq^{2n})^2}
\right),\\
\frac{~_B\langle 0;-|\sigma_1^- |-;0\rangle_B}{~_B\langle 0;-|-;0\rangle_B}
&=&0.
\end{eqnarray}
This is one of the main result of this paper.
Also,  sufficient conditions such that the
correlation functions for a triangular boundary coincide with those for a  diagonal boundary are derived. 
As a special case, we have the following equation for the diagonal matrix $\sigma^z$:
\begin{eqnarray}
\frac{~_B\langle i;\pm|\sigma_{M}^z\cdots\sigma_2^z \sigma_1^z |\pm;i\rangle_B}{
~_B\langle i;\pm|\pm;i\rangle_B}
=
\frac{~_B\langle i| \sigma_{M}^z\cdots\sigma_2^z \sigma_1^z |i\rangle_B}{
~_B\langle i|i\rangle_B}.
\end{eqnarray}

Finally, let us also mention that provided a suitable change of the boundary parameters, the Hamiltonian of the two triangular boundary models 
$H_B^{(\pm)}$ exchange each other under the action of the spin-reversal operator $\hat{\nu}$. As a consequence, correlation functions of the lower triangular model  $H_B^{(-)}$ are related to those of the upper triangular model $H_B^{(+)}$. Using this property, for instance we have the following identity of multiple integrals:
\begin{eqnarray}
&&
\frac{(q^4;q^4)_\infty^4}{(q^2;q^2)_\infty^8}
\frac{1-z^2}{\Theta_{q^4}(z^2)}
\left(2+\frac{1-z/r}{z}
\sum_{n=1}^\infty
(-q^2)^{n}\frac{
(z-z^{-1})-(1+q^{4n})/r+(z+z^{-1})q^{2n}
}{(1-q^{2n}z/r)(1-q^{2n}/rz)}\right)
\nonumber\\
&=&
\left(q^2 \int \int \int_{C_0}
-\int \int \int_{C_1}\right)
\prod_{a=1}^3 \frac{dw_a}{2\pi \sqrt{-1}}
\frac{
\displaystyle
q^2(1-1/rz)(1-q/rw_3)\prod_{a=1}^2(1-q^2/zw_a)}{
\displaystyle 
w_2^2 w_3^3 (1-q^2w_1/w_2)(1-q^4/w_1w_2)
\prod_{a=1}^2(1-q^2/rw_a)
}\nonumber\\
&\times&
\frac{
\Theta_{q^2}(w_1w_2)
\Theta_{q^2}(w_2/w_1)
\Theta_{q^2}(zw_3/q)
\Theta_{q^2}(qw_3/z)
\displaystyle
\prod_{a=1}^3
\Theta_{q^4}(w_a^2/q^2)}{
\displaystyle
\prod_{a=1}^2
\Theta_{q^2}(w_aw_3/q^2)
\Theta_{q^2}(w_a/qw_3)
\Theta_{q^2}(w_az)\Theta_{q^2}(w_a/z)},
\end{eqnarray}
where we have used the elliptic theta function
\begin{eqnarray}
\Theta_p(z)=(p;p)_\infty (z;p)_\infty (p/z;p)_\infty,~~~(z;p)_\infty=\prod_{n=0}^\infty(1-p^nz).
\label{def:theta}
\end{eqnarray}
Here the integration contours $C_l=C_l^{(+,1)}$ $(l=0,1)$ are simple closed curves 
given below (\ref{eqn:theta1}), (\ref{eqn:theta2}).

The plan of this paper is as follows.
In Section 2, the half-infinite $XXZ$ spin chain with a triangular boundary is formulated using the $q$-vertex operator approach.
In Section 3, we review the realizations of the vacuum eigenvectors and their duals \cite{JKKKM, BB1}.
In Section 4,  two integral representations of the correlation functions are calculated using bosonizations.
As a straightforward application, summation formulae of the boundary expectation values $\langle \sigma_1^\pm \rangle$ are obtained.
Also, we derive identities between multiple integrals of elliptic theta functions from
spin-reversal property.
 For each type of integral representation, a sufficient condition such that the expression for a triangular boundary condition coincides with
those for a diagonal boundary condition is identified.
Concluding remarks are given in Section 5. 
In Appendix \ref{appendix:A} we recall some basic facts about the quantum group $U_q(\widehat{sl_2})$
and fix the notations used in the main text.
In Appendix \ref{appendix:B} we recall the bosonizations of
$U_q(\widehat{sl_2})$ and the $q$-vertex operators.
In Appendix \ref{appendix:C} we summarize convenient formulae for the calculations of the vacuum expectation values.

\section{The half-infinite XXZ spin chain with a triangular boundary}

In this Section we give a mathematical formulation of the half-infinite
$XXZ$ spin chain with a triangular boundary, based on the $q$-vertex operator approach \cite{JKKKM}.

\subsection{Physical picture}

In this Section we sketch a physical picture of our problem. In Sklyanin's framework \cite{Sklyanin},  the transfer matrix 
$\widehat{T}_B^{(\pm,i)}(\zeta;r,s)$ that is a generating function of the Hamiltonian $H_B^{(\pm)}$ (\ref{def:Hamiltonian}) is introduced.
Basically, it is built from two objects: the  $R$-matrix and the $K-$matrix. For the model (\ref{def:Hamiltonian}), one introduces the $R$-matrix $R(\zeta)$ defined as:
\begin{eqnarray}
R(\zeta)=\frac{1}{\kappa(\zeta)}
\left(
\begin{array}{cccc}
1& & & \\
 &\frac{\displaystyle (1-\zeta^2)q}{\displaystyle 1-q^2\zeta^2}&\frac{
\displaystyle
(1-q^2)\zeta}{
\displaystyle
1-q^2\zeta^2}& \\
 &\frac{
\displaystyle
(1-q^2)\zeta}{
\displaystyle
1-q^2\zeta^2}&\frac{
\displaystyle
(1-\zeta^2)q}{
\displaystyle
1-q^2\zeta^2}& \\
 & & &1
\end{array}
\right),
\end{eqnarray}
where we have set
\begin{eqnarray}
\kappa(\zeta)=\zeta \frac{(q^4\zeta^2;q^4)_\infty (q^2/\zeta^2;q^4)_\infty}{
(q^4/\zeta^2;q^4)_\infty (q^2\zeta^2;q^4)_\infty},~~~
(z;p)_\infty=\prod_{n=0}^\infty (1-p^nz).\label{eq:kap}
\end{eqnarray}
Let $\{v_+,v_-\}$ denote the natural basis of $V={\bf C}^2$.
When viewed as an operator on $V \otimes V$,
the matrix elements of $R(\zeta) \in {\rm End}(V \otimes V)$ are given by
$R(\zeta)v_{\epsilon_1}\otimes v_{\epsilon_2}=
\sum_{\epsilon_1', \epsilon_2'=\pm} 
v_{\epsilon_1'}\otimes v_{\epsilon_2'}
R(\zeta)_{\epsilon_1' \epsilon_2'}^{\epsilon_1 \epsilon_2}$,
where the ordering of the index is given by
$v_+\otimes v_+, v_+\otimes v_-, v_-\otimes v_+, v_-\otimes v_-$.
As usual, when copies $V_j$ of $V$ are involved, $R_{i j}(\zeta)$ acts as
$R(\zeta)$ on the $i$-th and $j$-th components and as identity elsewhere.
The $R$-matrix $R(\zeta)$ satisfies the Yang-Baxter equation.
\begin{eqnarray}
R_{1 2}(\zeta_1/\zeta_2)
R_{1 3}(\zeta_1/\zeta_3)
R_{2 3}(\zeta_2/\zeta_3)
=
R_{2 3}(\zeta_2/\zeta_3)
R_{1 3}(\zeta_1/\zeta_3)
R_{1 2}(\zeta_1/\zeta_2).
\end{eqnarray}
The normalization factor $\kappa(\zeta)$
is determined by the following unitarity and crossing symmetry conditions:
\begin{eqnarray}
R_{12}(\zeta)R_{21}(\zeta^{-1})=1,~~~
R(\zeta)_{\epsilon_2 \epsilon_1'}^{\epsilon_2' \epsilon_1}=
R(-q^{-1}\zeta^{-1})_{-\epsilon_1 \epsilon_2}^{-\epsilon_1' \epsilon_2'}.
\end{eqnarray}
Also, we introduce the triangular $K$-matrix 
$K^{(\pm)}(\zeta)=K^{(\pm)}(\zeta;r,s)$ \cite{dVG, GZ} by
\begin{eqnarray}
K^{(+)}(\zeta;r,s)
&=&
\frac{\varphi(\zeta^2;r)}{\varphi(\zeta^{-2};r)}
\left(\begin{array}{cc}
\frac{\displaystyle 1-r\zeta^2}{\displaystyle \zeta^2-r}&
\frac{\displaystyle s \zeta(\zeta^2-\zeta^{-2})}{
\displaystyle \zeta^2-r}\\
0&1
\end{array}
\right),
\label{def:K+}\\
K^{(-)}(\zeta;r,s)
&=&
\frac{\varphi(\zeta^2;r)}{\varphi(\zeta^{-2};r)}
\left(\begin{array}{cc}
\frac{\displaystyle
1-r\zeta^2}{
\displaystyle
\zeta^2-r}&
0\\
\frac{
\displaystyle
s \zeta (\zeta^2-\zeta^{-2})}{
\displaystyle
\zeta^2-r}&1
\end{array}
\right),
\label{def:K-}
\end{eqnarray}
where we have set
\begin{eqnarray}
\varphi(z;r)=
\frac{(q^4rz;q^4)_\infty 
(q^6z^2;q^8)_\infty}{
(q^2 rz;q^4)_\infty 
(q^8z^2;q^8)_\infty}.
\end{eqnarray}
When viewed as an operator on $V$,
the matrix elements of $K^{(\pm)}(\zeta) \in {\rm End}(V)$ 
are given by
$K^{(\pm)}(\zeta)v_{\epsilon}=
\sum_{\epsilon'=\pm} 
v_{\epsilon'}
K^{(\pm)}(\zeta)_{\epsilon'}^{\epsilon}$,
where the ordering of the index is given by $v_+, v_-$.
As usual, when copies $V_j$ of $V$ are involved, $K_{j}^{(\pm)}(\zeta)$ acts as
$K^{(\pm)}(\zeta)$ on the $j$-th component and as identity elsewhere.
The $K$-matrix $K^{(\pm)}(\zeta)$ satisfies the boundary Yang-Baxter equation (also called the reflection equation):
\begin{eqnarray}
K_2^{(\pm)}(\zeta_2)
R_{21}(\zeta_1\zeta_2)
K_1^{(\pm)}(\zeta_1)
R_{12}(\zeta_1/\zeta_2)
=R_{21}(\zeta_1/\zeta_2)
K_1^{(\pm)}(\zeta_1)
R_{12}(\zeta_1\zeta_2)K_2^{(\pm)}(\zeta_2).
\label{eqn:BYBE}
\end{eqnarray}
The normalization factor (2.7) is determined by the following boundary unitarity and boundary crossing symmetry \cite{GZ}:
\begin{eqnarray}
K^{(\pm)}(\zeta)K^{(\pm)}(\zeta^{-1})=1,~~~
{K^{(\pm)}}(-q^{-1}\zeta^{-1})_{\epsilon_1}^{\epsilon_2}=
\sum_{\epsilon_1', \epsilon_2'=\pm}
R(-q\zeta^2)_{\epsilon_1' -\epsilon_2'}^{-\epsilon_1 \epsilon_2}
{K^{(\pm)}}(\zeta)_{\epsilon_2'}^{\epsilon_1'}.
\label{eqn:BUC}
\end{eqnarray}
The $K^{(\pm)}(\zeta;r,s)$ defined in (\ref{def:K+}) and (\ref{def:K-})
give general scalar triangular solutions of (\ref{eqn:BYBE}) and (\ref{eqn:BUC}).

In Sklyanin's framework, defined on a finite lattice the transfer matrix is built from a finite number of $R-$matrix \cite{Sklyanin}. In order to formulate the model (\ref{def:Hamiltonian}), an infinite combination of $R$-matrices \cite{JKKKM} is considered in $\widehat{T}_B^{(\pm,i)}(\zeta;r,s)$ . Generally speaking infinite combinations of the $R$-matrix are not free from the difficulty of divergence,
however we know two useful concepts to study infinite combinations of the $R$-matrix.
One is the corner transfer matrix (CTM) introduced by Baxter \cite{Baxter}.
The other is the $q$-vertex operator introduced by Baxter \cite{Baxter2} and Jimbo, Miwa, and Nakayashiki \cite{JMN}.
The CTM for $U_q(\widehat{sl_2})$ \cite{FM} gives
a supporting argument for the mathematical formulation presented in the next Section,
that is free from the difficulty of divergences.
 Following the strategy summarized in \cite{JKKKM, JMN}, let us recall the mathematical formulation of the $q$-vertex operators and the transfer matrix. Consider the infinite dimensional vector space  $\cdots \otimes V_3 \otimes V_2 \otimes V_1$   on which the Hamiltonian (1.1) acts.  Let us introduce the subspace ${\cal H}^{(i)}~(i=0,1)$
of the half-infinite spin chain by
\begin{eqnarray}
{\cal H}^{(i)}=Span \{
\cdots \otimes v_{p(N)} \otimes \cdots \otimes v_{p(2)} \otimes v_{p(1)}|~p(N)=(-1)^{N+i}~(N \gg 1)\},
\end{eqnarray}
where $p:{\bf N} \to \{\pm \}$.
We introduce 
the $q$-vertex operator $\widehat{\Phi}_\epsilon^{(1-i,i)}(\zeta)$
and the dual $q$-vertex operator $\widehat{\Phi}_\epsilon^{*(1-i,i)}(\zeta)$ for $\epsilon=\pm$
which act  on the space ${\cal H}^{(i)}$ $(i=0,1)$.
Their matrix elements are given by products of the $R$-matrix as follows:
\begin{eqnarray}
(\widehat{\Phi}_\epsilon^{(1-i,i)}(\zeta))
^{\cdots p(N)' \cdots p(2)' p(1)'}_{
\cdots p(N) \cdots p(2)~p(1)}
&=&
\lim_{N \to \infty}
\sum_{\mu(1), \mu(2), \cdots, \mu(N)=\pm}\prod_{j=1}^N R(\zeta)_{\mu(j-1)~p(j)}^{\mu(j)~p(j)'}
,\\
(\widehat{\Phi}_\epsilon^{*(1-i,i)}(\zeta))^{\cdots p(N)' \cdots p(2)' p(1)'}_{
\cdots p(N) \cdots p(2)~p(1)}&=&
\lim_{N \to \infty}
\sum_{\mu(1), \mu(2), \cdots, \mu(N)=\pm}\prod_{j=1}^N 
R(\zeta)_{p(j)~\mu(j)}^{p(j)'~\mu(j-1)},
\end{eqnarray}
where $\mu(0)=\epsilon$ and $\mu(N)=(-1)^{N+1-i}$.
We expect that the $q$-vertex operators $\widehat{\Phi}_\epsilon^{(1-i,i)}(\zeta)$ and
$\widehat{\Phi}_\epsilon^{*(1-i,i)}(\zeta)$ give rise to well-defined operators.
From heuristic arguments by using the $R$-matrix,
the $q$-vertex operators $\widehat{\Phi}_\epsilon^{(1-i,i)}(\zeta)$
and $\widehat{\Phi}_\epsilon^{* (1-i,i)}(\zeta)$ satisfy the following relations:
\begin{eqnarray}
\widehat{\Phi}_{\epsilon_2}^{(i,1-i)}(\zeta_2)
\widehat{\Phi}_{\epsilon_1}^{(1-i,i)}(\zeta_1)
&=&
\sum_{\epsilon_1', \epsilon_2'=\pm}R(\zeta_1/\zeta_2)_{\epsilon_1 \epsilon_2}^{\epsilon_1' \epsilon_2'}
\widehat{\Phi}_{\epsilon_1'}^{(i,1-i)}(\zeta_1)
\widehat{\Phi}_{\epsilon_2'}^{(1-i,i)}(\zeta_2),
\label{eqn:com-phys-VO}\\
\widehat{\Phi}_{\epsilon}^{*(1-i,i)}(\zeta)
&=&
\widehat{\Phi}_{-\epsilon}^{(1-i,i)}(-q^{-1}\zeta).
\label{eqn:dual-phys-VO}
\end{eqnarray}
From the property 
$R(\zeta)_{\epsilon_1, \epsilon_2}^{\epsilon_3, \epsilon_4}=
R(\zeta)_{-\epsilon_1, -\epsilon_2}^{-\epsilon_3, -\epsilon_4}$,
we have
\begin{eqnarray}
\hat{\nu}~\widehat{\Phi}_\epsilon^{(i,1-i)}(\zeta)~\hat{\nu}=
\widehat{\Phi}_{-\epsilon}^{(1-i,i)}(\zeta),
\label{eqn:reversal-phys-VO}
\end{eqnarray}
where we have used $\hat{\nu}=\prod_{j=1}^\infty \sigma_j^x$.
Following the strategy \cite{JKKKM} we introduce
the transfer matrix $\widehat{T}_B^{(\pm,i)}(\zeta; r,s)$ using the $q$-vertex operators.
\begin{eqnarray}
\widehat{T}_B^{(\pm,i)}(\zeta;r,s)=\sum_{\epsilon_1, \epsilon_2=\pm}
\widehat{\Phi}_{\epsilon_1}^{* (i,1-i)}(\zeta^{-1})K^{(\pm)}(\zeta;r,s)_{\epsilon_1}^{\epsilon_2}
\widehat{\Phi}_{\epsilon_2}^{(1-i,i)}(\zeta).
\label{def:phys-transfer}
\end{eqnarray}
From heuristic arguments by using the boundary Yang-Baxter equation 
(\ref{eqn:BYBE}) and relations of the $q$-vertex operators 
(\ref{eqn:com-phys-VO}) and (\ref{eqn:dual-phys-VO}), we have
the commutativity of the transfer matrix:
\begin{eqnarray}
~[\widehat{T}_B^{(\pm,i)}(\zeta_1;r,s),
\widehat{T}_B^{(\pm,i)}(\zeta_2;r,s)]=0~~~~{\rm for~any}~~\zeta_1, \zeta_2.
\end{eqnarray}
The Hamiltonian $H_B^{(\pm)}$ (\ref{def:Hamiltonian}) is obtained as
\begin{eqnarray}
\left.\frac{d}{d\zeta}\widehat{T}_B^{(\pm,i)}(\zeta;r,s)\right|_{\zeta=1}=\frac{4q}{1-q^2}H_B^{(\pm)}+{\rm const}.
\end{eqnarray}
In order to diagonalize the transfer matrix $\widehat{T}_B^{(\pm,i)}(\zeta;r,s)$, one follows the strategy that we call the $q$-vertex operator approach \cite{JKKKM}. As a guide for the structure of the eigenvectors of the Hamiltonian (\ref{def:Hamiltonian}), the observation that the $q$-Onsager algebra - a coideal subalgebra of $U_q(\widehat{sl_2})$ -  is the hidden symmetry of the Hamiltonian (\ref{def:Hamiltonian}) plays a central role.

It is important to stress that the formulation (\ref{def:phys-transfer}) independently arises within the so-called Onsager's approach
of the XXZ open spin chain \cite{BB1}. Indeed, the transfer matrix of the model (\ref{def:Hamiltonian}) can be formulated as the thermodynamic limit $N\rightarrow \infty$ of the transfer matrix of the finite model, which, in this framework, is expressed in terms of generators of the $q$-Onsager algebra \cite{BK}. In this limit, the transfer matrix associated with (\ref{def:Hamiltonian}) is a linear combination of $q$-Onsager currents. Either based on the central extension of the boundary Yang-Baxter equation \cite{BB2} or using the closed relationship between the $q$-Onsager algebra and a certain coideal subalgebra of $U_q(\widehat{sl_2})$ \cite{BK},  $q$-Onsager currents are realized as quadratic combinations of  $U_q(\widehat{sl_2})$  $q$-vertex operators, providing an alternative derivation of the transfer matrix (\ref{def:phys-transfer}).

\subsection{Vertex operator approach}

In this Section we give the mathematical formulation of our problem, the $q$-vertex operator approach to the half-infinite $XXZ$ spin chain with a triangular boundary.
Let $V_\zeta=V \otimes {\bf C}[\zeta, \zeta^{-1}]=V_\zeta^{(+)}\oplus V_\zeta^{(-)}$
where $V_\zeta^{(\pm)}={\rm span}\{v_\pm \otimes \zeta^{2n}, v_\mp \otimes \zeta^{2n-1}~(n \in {\bf Z})\}$. 
Let $V_\zeta$ the evaluation representation of $U_q(\widehat{sl_2})$ by setting
\begin{eqnarray}
&&
e_0 \cdot v_\epsilon \otimes \zeta^m=(f_1 v_\epsilon) \otimes \zeta^{m+1},~~
e_1 \cdot v_\epsilon \otimes \zeta^m=(e_1 v_\epsilon) \otimes \zeta^{m+1},\nonumber
\\
&&
f_0 \cdot v_\epsilon \otimes \zeta^m=(e_1 v_\epsilon) \otimes \zeta^{m-1},~~
f_1 \cdot v_\epsilon \otimes \zeta^m=(f_1 v_\epsilon) \otimes \zeta^{m-1},\\
&&
q^{h_0}=q^{-h_1},~~
q^{h_1} \cdot v_\epsilon \otimes \zeta^m=(q^{h_1}v_\epsilon)\otimes \zeta^m,~~
\rho=\zeta\frac{d}{d\zeta}\pm \frac{1}{2}~~{\rm on}~~V_\zeta^{(\pm)},
\nonumber
\end{eqnarray}
where the action of $e_1, f_1, q^{h_1}$ on $V$ are given by
$e_1 v_+=0$, $e_1v_-=v_+$, $f_1 v_+=v_-$, $f_1v_-=0$, and $q^{h_1}v_\pm=q^{\pm 1}v_\pm$.
Let $V(\Lambda_i)$
the irreducible highest weight $U_q(\widehat{sl_2})$ representation
with the fundamental weights $\Lambda_i$ $(i=0,1)$.
In other words,
there exists a vector $|i\rangle \in V(\Lambda_i)$ $(i=0,1)$ such that
\begin{eqnarray}
e_j |i\rangle=0,~q^{h_j}|i\rangle=q^{(h_j,\Lambda_i)}|i\rangle,
~f_j^{(h_j,\Lambda_i)+1}|i\rangle=0,
~V(\Lambda_i)=U_q(\widehat{sl_2})|i\rangle,
\end{eqnarray}
for $j=0,1$.
Let $V^*(\Lambda_i)$ the restricted dual representation of $V(\Lambda_i)$.
We introduce the type-I vertex operators $\Phi^{(1-i,i)}_\epsilon(\zeta)$
as the intertwiner of $U_q(\widehat{sl_2})$:
\begin{eqnarray}
\Phi^{(1-i,i)}(\zeta) : V(\Lambda_i) \longrightarrow V(\Lambda_{1-i}) \otimes V_\zeta,
&&
\Phi^{(1-i,i)}(\zeta) \cdot x=\Delta(x) \cdot \Phi^{(1-i,i)}(\zeta),\\
\Phi^{*(1-i,i)}(\zeta) : V(\Lambda_i) \otimes V_\zeta
\longrightarrow V(\Lambda_{1-i}),
&&
\Phi^{*(1-i,i)}(\zeta) \cdot \Delta(x)=x \cdot \Phi^{*(1-i,i)}(\zeta),
\end{eqnarray}
for $x \in U_q(\widehat{sl_2})$.
We set the elements of the type-I vertex operators
\begin{eqnarray}
\Phi^{(1-i,i)}(\zeta)=\sum_\epsilon \Phi_\epsilon^{(1-i,i)}(\zeta) \otimes v_\epsilon,~~~
\Phi^{*(1-i,i)}_\epsilon(\zeta)|v\rangle=\Phi^{*(1-i,i)}(\zeta)(|v\rangle \otimes v_\epsilon).
\end{eqnarray}
The type-I vertex operators
${\Phi}_\epsilon^{(1-i,i)}(\zeta)$ and
${\Phi}_\epsilon^{*(1-i,i)}(\zeta)$
satisfy the following commutation relation,
the duality, and the invertibility properties:
\begin{eqnarray}
&&
{\Phi}_{\epsilon_2}^{(i,1-i)}(\zeta_2)
{\Phi}_{\epsilon_1}^{(1-i,i)}(\zeta_1)
=
\sum_{\epsilon_1', \epsilon_2'=\pm}R(\zeta_1/\zeta_2)_{\epsilon_1 \epsilon_2}^{\epsilon_1' \epsilon_2'}
{\Phi}_{\epsilon_1'}^{(i,1-i)}(\zeta_1)
{\Phi}_{\epsilon_2'}^{(1-i,i)}(\zeta_2),
\label{eqn:com-math-VO}
\\
&&{\Phi}_{\epsilon}^{*(1-i,i)}(\zeta)
=
{\Phi}_{-\epsilon}^{(1-i,i)}(-q^{-1}\zeta),
\label{eqn:dual-math-VO}
\\
&&
g \sum_{\epsilon=\pm} \Phi^{* (i,1-i)}_\epsilon(\zeta)\Phi_\epsilon^{(1-i,i)}(\zeta)=id,~~~
g \Phi_{\epsilon_1}^{(i,1-i)}(\zeta)\Phi_{\epsilon_2}^{* (1-i,i))}(\zeta)=\delta_{\epsilon_1 \epsilon_2} id,
\label{eqn:inversion-math-VO}
\end{eqnarray}
where we have set
\begin{eqnarray}
g=\frac{(q^2;q^4)_\infty}{(q^4;q^4)_\infty}.
\end{eqnarray}
Following the strategy of \cite{JKKKM}, as the generating function of the Hamiltonian $H_B^{(\pm)}$ (\ref{def:Hamiltonian}) we introduce
the ``renormalized" transfer matrix ${T}_B^{(\pm,i)}(\zeta; r,s)$ using the $q$-vertex operators:
\begin{eqnarray}
{T}_B^{(\pm,i)}(\zeta;r,s)=g \sum_{\epsilon_1, \epsilon_2=\pm}
{\Phi}_{\epsilon_1}^{* (i,1-i)}(\zeta^{-1})K^{(\pm)}(\zeta;r,s)_{\epsilon_1}^{\epsilon_2}
{\Phi}_{\epsilon_2}^{(1-i,i)}(\zeta).
\label{def:math-transfer}
\end{eqnarray}
From the boundary Yang-Baxter equation and the properties of the $q$-vertex operators
(\ref{eqn:BYBE}), (\ref{eqn:BUC}), (\ref{eqn:com-math-VO}), 
(\ref{eqn:dual-math-VO}), (\ref{eqn:inversion-math-VO}), we have
the following properties of the ``renormalized" transfer matrix:
\begin{eqnarray}
&&~[{T}_B^{(\pm,i)}(\zeta_1;r,s),{T}_B^{(\pm,i)}(\zeta_2;r,s)]=0~~~~{\rm for~any}~~\zeta_1, \zeta_2,\\
&&~T_B^{(\pm,i)}(1;r,s)=id,~~~
T_B^{(\pm,i)}(\zeta;r,s)T_B^{(\pm,i)}(\zeta^{-1};r,s)=id.\\
&&~T_B^{(\pm,i)}(-q^{-1}\zeta^{-1};r,s)=T_B^{(\pm,i)}(\zeta;r,s).
\end{eqnarray}
Following the strategy in \cite{DFJMN, JKKKM, JMN} and the CTM argument for $U_q(\widehat{sl_2})$ \cite{FM}, as well as the alternative support within Onsager's framework \cite{BB1}, we study our problem upon the following identification:
\begin{eqnarray}
T_B^{(\pm,i)}(\zeta;r,s)=\widehat{T}_B^{(\pm,i)}(\zeta;r,s),
~~~\Phi_\epsilon^{(1-i,i)}(\zeta)=
\widehat{\Phi}_\epsilon^{(1-i,i)}(\zeta),
~~~\Phi_\epsilon^{*(1-i,i)}(\zeta)=
\widehat{\Phi}_\epsilon^{*(1-i,i)}(\zeta).
\end{eqnarray}
The point of using the $q$-vertex operators $\Phi_\epsilon^{(1-i,i)}(\zeta)$,  $\Phi_\epsilon^{*(1-i,i)}(\zeta)$
associated with $U_q(\widehat{sl_2})$ is that they are well-defined objects, free from the difficulty of divergence. In addition, they are the unique solution of the intertwining relations defining the $q$-vertex operators of the $q$-Onsager algebra (see \cite{BB1} for details), which characterizes the hidden non-Abelian symmetry of (\ref{def:Hamiltonian}). 

Finally, let us describe the spin-reversal property. 
Let $\nu : V(\Lambda_0) \longrightarrow V(\Lambda_1)$ be 
the vector-space isomorphism
corresponding to the Dynkin diagram symmetry \cite{DFJMN}. Then we have
\begin{eqnarray}
&&
\nu^{-1} e_j \nu=e_{1-j},~~
\nu^{-1} f_j \nu=f_{1-j},~~
\nu^{-1} q^{h_j} \nu=q^{h_{1-j}}.
\end{eqnarray}
Moreover we have
\begin{eqnarray}
\nu~\Phi_\epsilon^{(0,1)}(\zeta)~\nu=\Phi_{-\epsilon}^{(1,0)}(\zeta),
\label{eqn:reversal-VO}
\end{eqnarray}
which gives the same relation as (\ref{eqn:reversal-phys-VO}).
The $q$-vertex operator $\Phi_\epsilon^{(1-i,i)}(\zeta)$
has a bosonization summarized in appendix \ref{appendix:B}.
Note that we have two bosonizations of the $q$-vertex operator
based on (\ref{eqn:reversal-VO}).
Noting the relation
\begin{eqnarray}
\sigma^x K^{(\pm)}(\zeta;r,s) \sigma^x=
\Lambda(\zeta;r)
K^{(\mp)}(\zeta;1/r,-s/r),
\end{eqnarray}
where we have used
\begin{eqnarray}
\Lambda(\zeta;r)=\frac{1}{\zeta^2}\frac{\Theta_{q^4}(r\zeta^2)
\Theta_{q^4}(q^2r\zeta^{-2})}{
\Theta_{q^4}(r \zeta^{-2})
\Theta_{q^4}(q^2r \zeta^2)},
\end{eqnarray}
we find that
\begin{eqnarray}
\nu^{-1} T_B^{(\pm,1)}(\zeta;r,s) \nu=\Lambda(\zeta;r) T_B^{(\mp,0)}(\zeta;1/r,-s/r).
\label{eqn:reversal-transfer}
\end{eqnarray}
In the next Section we describe the vacuum eigenvectors $|\pm; i\rangle_B$  and their duals  $~_B\langle i;\pm|$  
 $(i=0,1)$ such that
\begin{eqnarray}
~_B\langle i;\pm|T_B^{(\pm, i)}(\zeta;r,s)
&=&
~_B\langle i;\pm|\Lambda^{(i)}(\zeta;r),
\label{def:dual-vacuum}\\
T_B^{(\pm, i)}(\zeta;r,s)|\pm; i\rangle_B
&=&
\Lambda^{(i)}(\zeta;r)|\pm ;i \rangle_B.
\label{def:vacuum}
\end{eqnarray}
Here we have set
\begin{eqnarray}
\Lambda^{(i)}(\zeta;r)=\left\{\begin{array}{cc}
1&~ (i=0)
\\
\Lambda(\zeta;r)&~(i=1)
\end{array}
\right..
\end{eqnarray}

Once the vacuum eigenvectors are found, it is possible to create
the excited states by an application of the type-II vertex operators.
Recall that type-II vertex operators $\Psi^{*(1-i,i)}_\epsilon(\xi)$
are the intertwiners of $U_q(\widehat{sl_2})$:
\begin{eqnarray}
\Psi^{*(1-i,i)}(\xi) : V_\xi \otimes V(\Lambda_i) \longrightarrow V(\Lambda_{1-i}),~~
\Psi^{*(1-i,i)}(\xi) \cdot \Delta(x)=x \cdot \Psi^{*(1-i,i)}(\xi),
\end{eqnarray}
for $x \in U_q(\widehat{sl_2})$.
Define the elements of the vertex operator
\begin{eqnarray}
\Psi_{\mu}^{*(1-i,i)}(\xi)|v\rangle=\Psi^{*(1-i,i)}(\xi)(v_{\mu} \otimes |v\rangle).
\end{eqnarray}
The type-I vertex operators
${\Phi}_\epsilon^{(1-i,i)}(\zeta)$ and
the type-II vertex operators
${\Psi}_\mu^{(i,1-i)}(\xi)$
satisfy the following commutation relation:
\begin{eqnarray}
{\Phi}_{\epsilon}^{(i,1-i)}(\zeta)
{\Psi}_{\mu}^{*(1-i,i)}(\xi)
&=&
\tau(\zeta/\xi)
{\Psi}_{\mu}^{*(i,1-i)}(\xi)
{\Phi}_{\epsilon}^{(1-i,i)}(\zeta),
\label{eqn:com-math-VO-I-II}
\end{eqnarray}
where we have set
\begin{eqnarray}
\tau(\zeta)=\zeta^{-1}\frac{\Theta_{q^4}(q\zeta^2)}{\Theta_{q^4}(q\zeta^{-2})}.
\end{eqnarray}
This commutation relation implies that 
\begin{eqnarray}
\Psi_{\mu_1}^{* (i+N,i+N-1)}(\xi_1)\cdots \Psi_{\mu_{N-1}}^{*(i+2,i+1)}(\xi_{N-1})\Psi_{\mu_N}^{*(i+1,i)}(\xi_N)
|\pm;i\rangle_B
\end{eqnarray}
is an eigenvector of $T_B^{(\pm,i)}(\zeta;r,s)$ with eigenvalue
\begin{eqnarray}
\Lambda^{(i)}(\zeta;r)\prod_{j=1}^N \tau(\zeta/\xi_j)\tau(\zeta \xi_j).
\end{eqnarray}
Here the number $i+N$ in the suffix $(i+N,i+N-1)$ of the $q$-vertex operator
$\Psi_{\mu_1}^{*(i+N,i+N-1)}(\xi_1)$ should be understood modulo $2$. Note that type-II vertex operators are intertwiners of the 
$q$-Onsager algebra, which justify their use in solving (\ref{def:Hamiltonian}) \cite{BB1}.

\section{Vacuum eigenvectors and their duals}

In the VOA, the computation of correlation functions essentially relies on the prior knowledge of realizations of the vacuum eigenvectors and their duals in terms of $q$-bosons. For the model (\ref{def:Hamiltonian}) with diagonal boundary conditions $s=0$, they have been proposed in [6]. For the model (1.1) with $s\neq 0$, vacuum eigenvectors have been proposed using the representation theory of the spectrum generating algebra ($q$-Onsager algebra). These objects  playing a central role in the analysis of further Sections, below we recall the basic notations and main results of \cite{JKKKM, BB1}. For completeness, dual vacuum eigenvectors  of the model (\ref{def:Hamiltonian}) are described in details.

\subsection{Diagonal boundary}

Upon the specialization $s=0$ the Hamiltonian $H_B^{(\pm)}$ degenerate to the one of
the $XXZ$ spin-chain with a diagonal boundary:
\begin{eqnarray}
H_B^{(\pm)}|_{s=0}=-\frac{1}{2}
\sum_{k=1}^\infty (\sigma_{k+1}^x \sigma_k^x+\sigma_{k+1}^y \sigma_k^y+\Delta \sigma_{k+1}^z \sigma_k^z)
-\frac{1-q^2}{4q}\frac{1+r}{1-r}\sigma_1^z.
\end{eqnarray}
Let us recall the structure of the the vacuum eigenvectors $|i\rangle_B$ and their duals
$~_B\langle i|$ for the $XXZ$ spin-chain with a diagonal boundary, obtained in \cite{JKKKM}. The spectral problem for the Hamiltonian is identical with the one for the ``renormalized" transfer matrix $T_B^{(\pm,i)}(\zeta;r,0)$, namely:
\begin{eqnarray}
T_B^{(\pm,i)}(\zeta;r,0)|i\rangle_B&=&\Lambda^{(i)}(\zeta;r) |i\rangle_B,
\label{def:vacuum-diagonal}\\
~_B\langle i|T_B^{(\pm,i)}(\zeta;r,0)&=&\Lambda^{(i)}(\zeta;r)_B\langle i|.
\label{def:dual-vacuum-diagonals}
\end{eqnarray}

Realizations of the vacuum eigenvectors and their duals follow from the bosonization of
the $q$-vertex operators. Indeed, recall that for  $i=0,1$ the bosonization of
the irreducible highest weight representation $V(\Lambda_i)$
and the restricted dual representation $V^*(\Lambda_i)$ with the fundamental weights $\Lambda_i$ is \cite{DFJMN} (see also Appendix \ref{appendix:B}):
\begin{eqnarray}
V(\Lambda_i)&=&{\bf C}[a_{-1}, a_{-2}, a_{-3}, \cdots]\otimes (\oplus_{n \in {\bf Z}}
{\bf C}e^{\Lambda_i+n \alpha}),
\label{eqn:bosonization-module}
\\
V^*(\Lambda_i)&=&{\bf C}[a_{1}, a_{2}, a_3, \cdots]\otimes (\oplus_{n \in {\bf Z}}
{\bf C}e^{-\Lambda_i-n \alpha}).
\label{eqn:bosonization-dual-module}
\end{eqnarray}
Note that the highest weight vector $|i\rangle$ and the lowest weight vector $\langle i|$ are given by \cite{DFJMN}:
\begin{eqnarray}
~\langle i|=1 \otimes e^{-\Lambda_i},~~~~~
|i\rangle=1\otimes e^{\Lambda_i}.
\end{eqnarray}
In \cite{JKKKM}, solving directly (\ref{def:dual-vacuum-diagonals}) using the bosonization of the transfer matrix that comes from
the one of the $q$-vertex operators (\ref{boson:VO1}) and (\ref{boson:VO2}), realizations of the vacuum eigenvectors and
their duals were obtained. The results of \cite{JKKKM} are summarized as follows:
\begin{eqnarray}
|i\rangle_B=\exp\left(F_i\right)|i\rangle, && F_i=\frac{1}{2}\sum_{n=1}^\infty
\frac{n \alpha_n}{[2n]_q[n]_q}a_{-n}^2+\sum_{n=1}^\infty \beta_n^{(i)}a_{-n},\label{def:vac}\\
~_B\langle i|=\langle i|\exp\left(G_i\right), && G_i=\frac{1}{2}\sum_{n=1}^\infty \frac{n \gamma_n}{[2n]_q[n]_q} a_n^2
+\sum_{n=1}^\infty \delta_n^{(i)}a_n.\label{def:vacdual}
\end{eqnarray}
Here we have set 
\begin{eqnarray}
&&
\alpha_n=-q^{6n},~~~\gamma_n=-q^{-2n},
\label{def:alpha}
\end{eqnarray}
and
\begin{eqnarray}
\beta_n^{(i)}&=&-\theta_n \frac{q^{{5n}/{2}}(1-q^n)}{[2n]_q}+\left\{
\begin{array}{cc}
-\frac{\displaystyle q^{{7n}/{2}}r^n}{\displaystyle [2n]_q}&~(i=0),\\
+\frac{\displaystyle q^{{3n}/{2}}r^{-n}}{\displaystyle [2n]_q}&~(i=1),
\end{array}
\right.\label{def:beta}
\\
\delta_n^{(i)}&=&
\theta_n \frac{q^{-{3n}/{2}}(1-q^n)}{[2n]_q}+\left\{
\begin{array}{cc}
-\frac{\displaystyle q^{-{5n}/{2}}r^n}{\displaystyle [2n]_q}&~(i=0),\\
+\frac{\displaystyle q^{-{n}/{2}}r^{-n}}{\displaystyle [2n]_q}&~(i=1).
\end{array}
\right.\label{def:delta}
\end{eqnarray}
where
\begin{eqnarray}
\theta_n=\left\{
\begin{array}{cc}
1& {\rm for}~n~{\rm even},\\
0& {\rm for}~n~{\rm odd}.
\end{array}
\right.
\end{eqnarray}
Note that the spin-reversal property of the ``renormalized" transfer matrix $T_B^{(\pm,i)}(\zeta;r,s)$ (\ref{eqn:reversal-transfer}) suggests
that the two vacuum eigenvectors $|i\rangle_B$ $(i=0,1)$ and their duals $~_B\langle i|$ $(i=0,1)$ 
should be related by
\begin{eqnarray}
\left.\nu\left(_B\langle 0|~\right)\right|_{r \to 1/r}=~_B\langle 1|,~~~\left.\nu\left(~|0\rangle_B~\right)\right|_{r \to 1/r}=|1\rangle_B.
\label{eqn:reversal-vacuum-diagonal}
\end{eqnarray}
However, since the spin-reversal symmetry is obscured in the bosonization,
we do not know how to verify this directly from the bosonization formulae.

\subsection{Triangular boundary}

In this Section we recall the structure of the vacuum eigenvectors and their duals
for the half-infinite $XXZ$ spin chain with a triangular boundary (\ref{def:Hamiltonian}).
For $s \in {\bf R}$ we are interested in the vacuum eigenvectors $|\pm;i\rangle_B$ which satisfy
\begin{eqnarray}
~T^{(\pm, i)}_B(\zeta;r,s)|\pm, i\rangle_B=\Lambda^{(i)}(\zeta;r)|\pm;i\rangle_B.
\label{def:vacuum-triangular}
\end{eqnarray}
Clearly, the Hamiltonian (\ref{def:Hamiltonian}) can be considered as an integrable perturbation of the diagonal boundary case $s=0$. In addition, the spectrum generating algebra associated with (\ref{def:Hamiltonian}) is the $q$-Onsager algebra \cite{BB1}. As a consequence, any eigenvector of the transfer matrix $T^{(\pm, i)}_B(\zeta;r,s)$ associated with the Hamiltonian (\ref{def:Hamiltonian})
can be potentially written in terms of monomials of the $q$-Onsager generators acting on $|i\rangle_B$. In the model (\ref{def:Hamiltonian}), recall that realizations of the $q$-Onsager fundamental generators are known in terms of $U_q(\widehat{sl_2})$ Drinfeld's basic generators \cite{BK}. For the vacuum eigenvectors $|\pm;i\rangle_B$, it is thus natural to look for a combinations of monomials in terms of basic Drinfeld generators
acting on $|i\rangle_B$. The results of \cite{BB1} are summarized as follows:
\begin{eqnarray}
&&|+;0\rangle_B =\exp_{q^{-1}}\left(s f_0 \right)|0\rangle_B,
\\
&&|+;1\rangle_B=\exp_{q}\left(\frac{~s~}{rq}e_1 q^{-h_1}\right)|1\rangle_B,
\\
&&
|-;0\rangle_B =\exp_{q}\left(-\frac{~s~}{q}e_0 q^{-h_0}\right)|0\rangle_B,
\\
&&
|-;1\rangle_B=\exp_{q^{-1}}\left(-\frac{~s~}{r}f_1 \right)|1\rangle_B,
\end{eqnarray}
where $|i\rangle_B$ are the vacuum eigenvectors (\ref{def:vac}) of the diagonal boundary $s=0$ \cite{JKKKM}.
Here we have used the $q$-exponential function
$\exp_q(x)$ given in (\ref{def:q-exp}).
In order to show the relation (\ref{def:vacuum-triangular}),  the following intertwining property of the $q$-vertex operators are needed:
\begin{eqnarray}
\Phi_-^{(1-i,i)}(\zeta){\cal B}_{+,0}^n &=&
q^{-n} {\cal B}_{+,0}^n \Phi_-^{(1-i,i)}(\zeta),\\
\Phi_+^{(1-i,i)}(\zeta){\cal B}_{+,0}^n&=&q^n {\cal B}_{+,0}^n \Phi_+^{(1-i,i)}(\zeta)
+s~\zeta^{-1} [n]_q {\cal B}_{+,0}^{n-1}
\Phi_-^{(1-i,i)}(\zeta),\\
\Phi_-^{(1-i,i)}(\zeta){\cal B}_{+,1}^n &=&
q^{n} {\cal B}_{+,1}^n \Phi_-^{(1-i,i)}(\zeta),\\
\Phi_+^{(1-i,i)}(\zeta)
{\cal B}_{+,1}^n&=&q^{-n} {\cal B}_{+,1}^n \Phi_+^{(1-i,i)}(\zeta)
+\frac{~s~}{r} \zeta [n]_q {\cal B}_{+,1}^{n-1}\Phi_-^{(1-i,i)}(\zeta),
\\
\Phi_-^{(1-i,i)}(\zeta){\cal B}_{-,0}^n&=&
q^{-n}{\cal B}_{-,0}^n\Phi_-^{(1-i,i)}(\zeta)+
s~\zeta [n]_q {\cal B}_{-,0}^{n-1}\Phi_+^{(1-i,i)}(\zeta),
\\
\Phi_+^{(1-i,i)}(\zeta){\cal B}_{-,0}^n&=&
q^n {\cal B}_{-,0}^n \Phi_+^{(1-i,i)}(\zeta),
\\
\Phi_-^{(1-i,i)}(\zeta){\cal B}_{-,1}^n
&=&q^{n}{\cal B}_{-,1}^n\Phi_-^{(1-i,i)}(\zeta)
+\frac{~s~}{r}\zeta^{-1} [n]_q {\cal B}_{-,1}^{n-1}\Phi_+^{(1-i,i)}(\zeta),\\
\Phi_+^{(1-i,i)}(\zeta){\cal B}_{-,1}^n&=&q^{-n} {\cal B}_{-,1}^n
\Phi_+^{(1-i,i)}(\zeta).
\end{eqnarray}
Here we have set
\begin{eqnarray}
{\cal B}_{+,0}=s f_0,~~
{\cal B}_{+,1}=\frac{~s~}{r q}e_1 q^{-h_1},~~
{\cal B}_{-,0}=\frac{~s~}{q}e_0 q^{-h_0},~~
{\cal B}_{-,1}=\frac{~s~}{r}f_1.
\end{eqnarray}
Once we assume the relation (\ref{eqn:reversal-vacuum-diagonal}),  the spin-reversal property of the vacuum eigenvectors follows directly from the $q$-exponential formula. 	Indeed,
\begin{eqnarray}
\left.\nu\left(|\pm; 0 \rangle_B\right)\right|_{r \to 1/r, s\to -s/r}=|\mp; 1\rangle_B,
\label{eqn:reversal-vacuum-triangular}
\end{eqnarray}
which is consequence of the following reversal property
\begin{eqnarray}
\left.\nu^{-1} {\cal B}_{\pm,1}\nu \right|_{r\to 1/r, s\to-s/r}=-{\cal B}_{\mp,0}.
\end{eqnarray}

For $s \in {\bf R}$,  we are now interested in the dual vacuum eigenvectors given by
\begin{eqnarray}
~_B\langle i;\pm|T^{(\pm,i)}_B(\zeta;r,s)=\Lambda^{(i)}(\zeta;r)~_B\langle i;\pm|.
\label{def:dual-vacuum-triangular}
\end{eqnarray}
Above arguments also hold for the construction of the dual vacuum eigenvectors, from which we obtain the following  realizations:
\begin{eqnarray}
&&
~_B\langle 0; +|=~_B\langle 0|\exp_q\left(-s f_0\right),
\\
&&
~_B\langle 1;+|=~_B\langle 1|\exp_{q^{-1}}\left(-\frac{~s~}{r q} e_1 q^{-h_1}\right),
\\
&&~_B\langle 0; -|=~_B\langle 0|\exp_{q^{-1}}\left(\frac{~s~}{q} e_0 q^{-h_0}\right),
\\
&&~_B\langle 1;-|=~_B\langle 1|\exp_{q}\left(\frac{~s~}{r} f_1\right).
\end{eqnarray}
Here $~_B\langle i|$ are the dual vacuum eigenvectors (\ref{def:vacdual}) of the diagonal boundary $s=0$ \cite{JKKKM}.
In order to show the relation (\ref{def:dual-vacuum-triangular}), 
we need the following intertwining property of the dual $q$-vertex operators:
\begin{eqnarray}
\Phi_+^{* (1-i,i)}(\zeta){\cal B}_{+,0}^n &=&
q^{-n} {\cal B}_{+,0}^n \Phi_+^{* (1-i,i)}(\zeta),\\
\Phi_-^{* (1-i,i)}(\zeta){\cal B}_{+,0}^n&=&q^n {\cal B}_{+,0}^n 
\Phi_-^{* (1-i,i)}(\zeta)-q s \zeta^{-1} [n]_q {\cal B}_{+,0}^{n-1}
\Phi_+^{* (1-i,i)}(\zeta),\\
\Phi_+^{* (1-i,i)}(\zeta){\cal B}_{+,1}^n &=&
q^{n} {\cal B}_{+,1}^n \Phi_+^{* (1-i,i)}(\zeta),\\
\Phi_-^{* (1-i,i)}(\zeta){\cal B}_{+,1}^n&=&q^{-n} {\cal B}_{+,1}^n 
\Phi_-^{* (1-i,i)}(\zeta)-\frac{~s~}{r q} \zeta [n]_q {\cal B}_{+,1}^{n-1}
\Phi_+^{* (1-i,i)}(\zeta),\\
\Phi_+^{* (1-i,i)}(\zeta){\cal B}_{-,0}^n&=&
q^{-n}{\cal B}_{-,0}^n\Phi_+^{* (1-i,i)}(\zeta)-\frac{~s~}{q}\zeta [n]_q {\cal B}_{-,0}^{n-1}
\Phi_-^{* (1-i,i)}(\zeta),\\
\Phi_-^{* (1-i,i)}(\zeta){\cal B}_{-,0}^n&=&q^n {\cal B}_{-,0}^n \Phi_-^{* (1-i,i)}(\zeta),
\\
\Phi_+^{* (1-i,i)}(\zeta){\cal B}_{-,1}^n&=&q^{n}{\cal B}_{-,1}^n
\Phi_+^{* (1-i,i)}(\zeta)
-\frac{q s}{r}\zeta^{-1} [n]_q {\cal B}_{-,1}^{n-1}\Phi_-^{* (1-i,i)}(\zeta),\\
\Phi_-^{* (1-i,i)}(\zeta){\cal B}_{-,1}^n&=&q^{-n} {\cal B}_{-,1}^n\Phi_-^{* (1-i,i)}(\zeta).
\end{eqnarray}
Once we assume the relation (\ref{eqn:reversal-vacuum-diagonal}), the spin-reversal property of the dual vacuum eigenvectors
for a triangular boundary condition follows directly from the $q$-exponential formula:
\begin{eqnarray}
\left.\nu\left(_B\langle 0; \pm |\right)\right|_{r \to 1/r, s\to -s/r}=~_B\langle 1; \mp|.
\label{eqn:reversal-dual-vacuum-triangular}
\end{eqnarray}

Finally, from the relation for the $q$-exponential function $\exp_q(x)\exp_{q^{-1}}(-x)=1$ we deduce
\begin{eqnarray}
~_B\langle i; \pm|\pm; i\rangle_B=~_B\langle i|i\rangle_B.
\end{eqnarray}
Hence we have
\begin{eqnarray}
~_B\langle i; \pm|\pm; i\rangle_B=\left\{
\begin{array}{cc}
\frac{\displaystyle
(q^4r^2;q^8)_\infty}{
\displaystyle
(q^6;q^8)_\infty (q^2r^2;q^8)_\infty}&~~(i=0),\\
\frac{
\displaystyle
(q^4/r^2;q^8)_\infty}{
\displaystyle
(q^6;q^8)_\infty (q^2/r^2;q^8)_\infty}&~~(i=1).
\end{array}\right.
\end{eqnarray}
Here we have used the formulae of the norms of 
$~_B\langle i|i\rangle_B$ given in \cite{JKKKM}.

\section{Correlation functions}

In this Section, two integral representations for correlation functions of $q$-vertex operators are proposed. In particular, it is shown that the expressions obtained for a subset of correlation functions for the triangular boundary case coincide with the ones associated with a diagonal boundary. Based on the exact relation between local spin operators and $q$-vertex operators \cite{DFJMN,JKKKM}, summation formulae for the boundary expectation value of the spin operator in the models (\ref{def:Hamiltonian}) are derived.  In the last subsection, using the spin-reversal property we deduce linear relations between certain multiple integrals involving elliptic theta functions. The simplest examples are presented.

\subsection{Definitions}

In this Section, we focus our attention on
the vacuum expectation values of products of the $q$-vertex operators given by
\begin{eqnarray}
P^{(\pm,i)}_{\epsilon_1, \epsilon_2, \cdots,\epsilon_M}(\zeta_1, \zeta_2, \cdots,\zeta_M;r,s)=
\frac{~_B\langle i;\pm|\Phi_{\epsilon_1}^{(i,1-i)}(\zeta_1)
\Phi_{\epsilon_2}^{(1-i,i)}(\zeta_2)
\cdots
\Phi_{\epsilon_{M}}^{(1-i,i)}(\zeta_{M})|\pm ; i\rangle_B
}{~_B\langle i;\pm|\pm; i \rangle_B},
\label{def:M-point}
\end{eqnarray}
with $M$ an even integer. Our purpose is to derive them as integrals of meromorphic functions involving
infinite products (\ref{eqn:correlation1}) and (\ref{eqn:correlation2}), which will be detailed in the next two subsections.
In particular, as we will show in the next Section, upon the condition $\sum_{j=1}^M \epsilon_j=0$,
the vacuum expectation value of triangular boundary coincides with the one of diagonal boundary. Namely,
\begin{eqnarray}
\frac{~_B\langle i;\pm|\Phi_{\epsilon_1}^{(i,1-i)}(\zeta_1)
\cdots
\Phi_{\epsilon_{M}}^{(1-i,i)}(\zeta_{M})|\pm ; i\rangle_B
}{~_B\langle i;\pm|\pm; i \rangle_B}=
\frac{~_B\langle i|\Phi_{\epsilon_1}^{(i,1-i)}(\zeta_1)
\cdots
\Phi_{\epsilon_{M}}^{(1-i,i)}(\zeta_{M})| i\rangle_B
}{~_B\langle i | i \rangle_B}.
\end{eqnarray}

Note that from the spin-reversal properties 
(\ref{eqn:reversal-VO}), (\ref{eqn:reversal-vacuum-triangular}), (\ref{eqn:reversal-dual-vacuum-triangular}),
we have
\begin{eqnarray}
P^{(\pm,i)}_{\epsilon_1,\cdots,\epsilon_M}(\zeta_1,\cdots,\zeta_M;r,s)=
P^{(\mp,1-i)}_{-\epsilon_1,\cdots,-\epsilon_M}(\zeta_1,\cdots,\zeta_M;1/r,-s/r),
\label{eqn:reversal-correlation}
\end{eqnarray}
which will be used in the last subsection.
We have two bosonizations of the $q$-vertex operator,
which are based on the relation : $\nu \Phi_\epsilon^{(0,1)}(\zeta) \nu
=\Phi_{-\epsilon}^{(1,0)}(\zeta)$.
Hence we have two formulae of the correlation functions
in (\ref{eqn:reversal-correlation}).

Upon specialization of the spectral parameters (see \cite{DFJMN,JKKKM}),
the vacuum expectation values (\ref{def:M-point}) give multi-point correlation functions of local spin operators of 
the half-infinite $XXZ$ spin chain with a triangular boundary (\ref{def:Hamiltonian}).
Let $L$ be a linear operator on the $n$-fold tensor products
of the two-dimensional space $V_n \otimes \cdots \otimes V_2 \otimes V_1$.
The corresponding local operator ${\cal L}$ acting on $V(\Lambda_i)$
can be defined in terms of the type-I vertex operators in exactly the same way as in the bulk theory \cite{DFJMN}.
Explicitly, if $L$ is the matrix at the $n$-th site
\begin{eqnarray}
E_{\epsilon \epsilon'}\otimes \underbrace{id \otimes \cdots \otimes id}_{n-1},
\end{eqnarray}
the corresponding local operator ${\cal E}_{\epsilon \epsilon'}$
is given by
\begin{eqnarray}
{\cal E}_{\epsilon \epsilon'}=E_{\epsilon \epsilon'}(1,1,\cdots,1),
\end{eqnarray}
where we have set
\begin{eqnarray}
E_{\epsilon \epsilon'}(\zeta_1,\zeta_2,\cdots,\zeta_n)
&=&
g^n
\sum_{\epsilon_1,\cdots,\epsilon_{n-1}=\pm}
\Phi_{\epsilon_1}^{* (i,i+1)}(\zeta_1)
\cdots 
\Phi_{\epsilon_{n-1}}^{*(i+n-2,i+n-1)}(\zeta_{n-1})
\Phi_{\epsilon}^{* (i+n-1,i+n)}(\zeta_n)\nonumber\\
&\times&
\Phi_{\epsilon'}^{(i+n,i+n-1)}(\zeta_n)
\Phi_{\epsilon_{n-1}}^{(i+n-2,i+n-1)}(\zeta_{n-1})
\cdots
\Phi_{\epsilon_1}^{(i+1,i)}(\zeta_1).
\end{eqnarray}
From the inversion property of the $q$-vertex operators (\ref{eqn:inversion-math-VO}),
we have
\begin{eqnarray}
&&\frac{\displaystyle
~_B\langle i;\pm|E_{\epsilon_n \epsilon_n'}(\zeta_1,\cdots,\zeta_n)
E_{\epsilon_{n-1} \epsilon_{n-1}'}(\zeta_1,\cdots,\zeta_{n-1})
\cdots
E_{\epsilon_1 \epsilon_1'}(\zeta_1)|\pm;i\rangle_B}
{~_B\langle i;\pm|\pm;i \rangle_B}\nonumber\\
&=&
g^n P_{-\epsilon_1,-\epsilon_2,\cdots,-\epsilon_n,\epsilon_n',\cdots,\epsilon_2',\epsilon_1'}^{(\pm,i)}
(-q^{-1}\zeta_1,-q^{-1}\zeta_2,\cdots,-q^{-1}\zeta_n,\zeta_n,\cdots,\zeta_2,\zeta_1).
\end{eqnarray}
As a consequence, correlation functions of local operators are given by:
\begin{eqnarray}
&&
\frac{~_B\langle i;\pm|{\cal E}_{\epsilon_n \epsilon_n'}
{\cal E}_{\epsilon_{n-1} \epsilon_{n-1}'}
\cdots
{\cal E}_{\epsilon_1 \epsilon_1'}
|\pm;i\rangle_B
}{~_B\langle i;\pm|\pm;i\rangle_B}
= g^n P_{-\epsilon_1,\cdots,-\epsilon_n,\epsilon_n',\cdots,\epsilon_1'}^{(\pm,i)}
(-q^{-1},\cdots,-q^{-1}, 1, \cdots, 1).
\end{eqnarray}
In what follows we calculate the vacuum expectation values explicitly using bosonizations.

\subsection{First integral representation}

Consider the  $M$-point functions with $M$ an even integer :
\begin{eqnarray}
P^{(+,i)}_{\epsilon_1 \cdots \epsilon_M}(\zeta_1,\cdots,\zeta_M;r,s)=
P^{(-,1-i)}_{-\epsilon_1 \cdots -\epsilon_M}(\zeta_1,\cdots,\zeta_M;1/r,-s/r).
\end{eqnarray}
Here we calculate 
the vacuum expectation value using bosonizations
associated with the upper-triangular boundary model $H_B^{(+)}$.
In what follows, it is convenient to set $z_j=\zeta_j^2$ and
\begin{eqnarray}
A=\{1\leq a \leq M|\epsilon_a=+\}.\label{def:A}
\end{eqnarray}
The normal ordering of products of the type-I vertex operators are given by
\begin{eqnarray}
&&\Phi_{\epsilon_1}^{(i,1-i)}(\zeta_1)\cdots \Phi_{\epsilon_M}^{(1-i,i)}(\zeta_{M})\nonumber\\
&=&
(-q^3)^{\frac{1}{4}M(M-1)-\sum_{a \in A}a}(1-q^2)^{|A|}
\prod_{j=1}^{M} \zeta_j^{\frac{1+\epsilon_j}{2}+M-j} \prod_{1\leq j<k \leq M}
\frac{(q^2z_k/z_j;q^4)_\infty}{
(q^4z_k/z_j;q^4)_\infty}
\nonumber
\\
&\times&
\prod_{a \in A}\oint_C \frac{dw_a}{2\pi \sqrt{-1}} w_a
\frac{\displaystyle \prod_{a,b \in A \atop{a<b}}(w_a-w_b)(w_a-q^2w_b)}{
\displaystyle
\prod_{a \in A}
\left\{
\prod_{1 \leq j \leq a}(z_j-q^{-2}w_a)
\prod_{a \leq j \leq M}(w_a-q^4 z_j)
\right\}}\nonumber\\
&\times&:
\Phi_{-}^{(i,1-i)}(\zeta_1)\cdots \Phi_{-}^{(1-i,i)}(\zeta_{M})
\prod_{a \in A}X^-(w_a):.
\end{eqnarray}
Here the integration contour $C$ encircles $w_a=0~(a \in A)$ in such a way
that
$q^4z_j~(a\leq j \leq M)$ is inside and $q^2z_j~(1\leq j \leq a)$ is outside.
From the bosonizations of the Drinfeld's realization in Appendix \ref{appendix:B}, we have
\begin{eqnarray}
f_0=q^{h_1}x_{-1}^+=q^{\partial} \oint \frac{dv}{2\pi \sqrt{-1} v}X^+(v),~~~ 
e_1 q^{-h_1}=x_0^+ q^{-h_1}=
\oint \frac{dv}{2\pi \sqrt{-1}}X^+(v) q^{-\partial}.
\end{eqnarray}
From the normal orderings in appendix \ref{appendix:B} we have
\begin{eqnarray}
(f_0)^l&=&q^{l(l+1)}\prod_{b=1}^{l} \oint \frac{dv_b}{2\pi \sqrt{-1} v_b}
\prod_{1 \leq a<b \leq l}(v_a-v_b)(v_a-q^{-2}v_b)
:\prod_{b=1}^l X^+(v_b):q^{l \partial},
\\
(e_1 q^{-h_1})^l
&=&
q^{-l(l-1)}
\prod_{b=1}^l \oint \frac{d v_b}{2\pi \sqrt{-1}}
\prod_{1 \leq a<b \leq l}(v_a-v_b)(v_a-q^{-2}v_b)
:\prod_{b=1}^l X^+(v_b):q^{-l \partial}.
\end{eqnarray}
Hence we have the following normal orderings
\begin{eqnarray}
&&
\exp_q\left(-s f_0\right) \cdot \Phi_{\epsilon_1}^{(0,1)}(\zeta_1)
\Phi_{\epsilon_2}^{(1,0)}(\zeta_2) \cdots \Phi_{\epsilon_{M}}^{(1,0)}(\zeta_{M}) 
\cdot \exp_{q^{-1}}\left(s f_0\right)\nonumber\\
&=&
(-q^3)^{\frac{1}{4}M(M-1)-\sum_{a \in A}a}
(1-q^2)^{|A|}\prod_{j=1}^{M} 
\zeta_j^{\frac{1+\epsilon_j}{2}+M-j}\prod_{1\leq j<k \leq M}
\frac{(q^2z_k/z_j;q^4)_\infty}{(q^4z_k/z_j;q^4)_\infty}
\nonumber\\
&\times&
\sum_{n=0}^\infty s^n
\left\{\sum_{l,m \geq 0
\atop{l+m=n}}
\frac{(-1)^l }{[l]_q![m]_q!}q^{
\frac{3}{2}l^2+\frac{l}{2}+\frac{1}{2}m^2+\frac{3}{2}m+
l(2m+M-2|A|)}\right\}\nonumber\\
&\times&
\prod_{b=1}^n \oint \frac{dv_b}{2\pi \sqrt{-1} v_b}
\prod_{a \in A}\oint \frac{dw_a}{2\pi \sqrt{-1}}w_a
\frac{\displaystyle \prod_{b=1}^n \prod_{j=1}^M (v_b-q^3z_j)}{
\displaystyle \prod_{a \in A}
\left\{
\prod_{j \leq a}
(z_j-q^{-2}w_a)
\prod_{a \leq j}
(w_a-q^4z_j)
\right\}}\nonumber\\
&\times&
\frac{\displaystyle
\prod_{1\leq a<b \leq n}
(v_a-v_b)(v_a-q^{-2}v_b)
\prod_{a,b \in A
\atop{a<b}}(w_a-w_b)(w_a-q^2w_b)
}{
\displaystyle
\prod_{b=1}^n
\prod_{a \in A}(v_b-qw_a)(v_b-q^{-1}w_a)}\nonumber\\
&\times&
:\Phi_{-}^{(0,1)}(\zeta_1)\Phi_-^{(1,0)}(\zeta_2)
\cdots 
\Phi_-^{(1,0)}(\zeta_{M})\prod_{a \in A}X^-(w_a)
\prod_{b=1}^n X^+(v_b): q^{n \partial},
\end{eqnarray}
and
\begin{eqnarray}
&&
\exp_{q^{-1}}\left(-\frac{~s~}{r q} e_1q^{-h_1}\right) \cdot \Phi_{\epsilon_1}^{(1,0)}(\zeta_1)
\Phi_{\epsilon_2}^{(0,1)}(\zeta_2) \cdots \Phi_{\epsilon_M}^{(0,1)}(\zeta_M) \cdot 
\exp_{q}\left(\frac{~s~}{r q} e_1 q^{-h_1}\right)\nonumber\\
&=&
(-q^3)^{\frac{1}{4}M(M-1)-\sum_{a \in A}a}
(1-q^2)^{|A|}\prod_{j=1}^{M} 
\zeta_j^{\frac{1+\epsilon_j}{2}+M-j}\prod_{1\leq j<k \leq M}
\frac{(q^2z_k/z_j;q^4)_\infty}{(q^4z_k/z_j;q^4)_\infty}
\nonumber\\
&\times&
\sum_{n=0}^\infty (s/rq)^n
\left\{\sum_{l,m \geq 0
\atop{l+m=n}}
\frac{(-1)^l }{[l]_q![m]_q!}q^{
-\frac{3}{2}l(l-1)-\frac{1}{2}m(m-1)-l(2m+M-2|A|)}\right\}\nonumber\\
&\times&
\prod_{b=1}^n \oint \frac{dv_b}{2\pi \sqrt{-1}}
\prod_{a \in A}\oint \frac{dw_a}{2\pi \sqrt{-1}}w_a
\frac{\displaystyle \prod_{b=1}^n \prod_{j=1}^M (v_b-q^3z_j)}{
\displaystyle \prod_{a \in A}
\left\{
\prod_{j \leq a}
(z_j-q^{-2}w_a)
\prod_{a \leq j}
(w_a-q^4z_j)
\right\}}\nonumber\\
&\times&
\frac{\displaystyle
\prod_{1\leq a<b \leq n}
(v_a-v_b)(v_a-q^{-2}v_b)
\prod_{a,b \in A
\atop{a<b}}(w_a-w_b)(w_a-q^2w_b)
}{
\displaystyle
\prod_{b=1}^n
\prod_{a \in A}(v_b-qw_a)(v_b-q^{-1}w_a)}\nonumber\\
&\times&
:\Phi_{-}^{(1,0)}(\zeta_1)\Phi_-^{(0,1)}(\zeta_2)
\cdots 
\Phi_-^{(0,1)}(\zeta_{M})\prod_{a \in A}X^-(w_a)
\prod_{b=1}^n X^+(v_b): q^{-n \partial}.
\end{eqnarray}
The zero-mode $e^{\alpha}$ part of the operator 
$
:\Phi_{-}^{(i,1-i)}(\zeta_1)\Phi_-^{(1-i,i)}(\zeta_2)
\cdots 
\Phi_-^{(1-i,i)}(\zeta_{M})\prod_{a \in A}X^-(w_a)
\prod_{b=1}^n X^+(v_b):
q^{(1-2i)n \partial}
$
is given by 
$e^{(n-|A|+\frac{M}{2})\alpha}$.
Then, the condition for which the vacuum expectation value is non-vanishing reads:
\begin{eqnarray}
&&~_B\langle i|
:\Phi_{-}^{(i,1-i)}(\zeta_1)
\cdots 
\Phi_-^{(1-i,i)}(\zeta_{M})\prod_{a \in A}X^-(w_a)
\prod_{b=1}^n X^+(v_b):
q^{(1-2i)n \partial}
|i\rangle_B \neq 0,\nonumber\\
&&
\Longleftrightarrow 
~~_B\langle i|e^{(n-|A|+\frac{M}{2})\alpha}|i\rangle_B\neq 0~~
\Longleftrightarrow~~|A|-\frac{M}{2}=n \geq 0.
\label{prop:operator-zero-1}
\end{eqnarray}

Hence, provided the condition $|A| \geq \frac{M}{2}$ is satisfied, non-zero vacuum expectation values of type-I vertex operators are given by:
\begin{eqnarray}
&&
~_B\langle i;+|\Phi_{\epsilon_1}^{(i,1-i)}(\zeta_1)
\Phi_{\epsilon_2}^{(1-i,i)}(\zeta_2)
\cdots
\Phi_{\epsilon_{M}}^{(1-i,i)}(\zeta_{M})|+;i \rangle_B\nonumber\\
&=&(qs)^{|A|-\frac{M}{2}}
(-q^2 r)^{(\frac{M}{2}-|A|)i}(-q^3)^{\frac{1}{4}M^2+\frac{1}{2}M i -\sum_{a \in A}a}(1-q^2)^{|A|}
\nonumber\\
&\times&
\left\{\sum_{l,m \geq 0
\atop{l+m=|A|-\frac{M}{2}}}
\frac{(-1)^l }{[l]_q![m]_q!}q^{
-\frac{1}{2}l(l+1)+\frac{1}{2}m(m+1)}\right\}
\prod_{j=1}^{M} 
\zeta_j^{\frac{1+\epsilon_j}{2}+M-j+i}\prod_{1\leq j<k \leq M}
\frac{(q^2z_k/z_j;q^4)_\infty}{(q^4z_k/z_j;q^4)_\infty}
\nonumber\\
&\times&
\prod_{b=1}^{|A|-\frac{M}{2}} 
\oint \frac{dv_b}{2\pi \sqrt{-1}}v_b^{2i-1}
\prod_{a \in A}
\oint \frac{dw_a}{2\pi \sqrt{-1}}w_a^{1-i}
\frac{\displaystyle 
\prod_{b=1}^{|A|-\frac{M}{2}} \prod_{j=1}^{M}
 (v_b-q^3z_j)}{
\displaystyle \prod_{a \in A}
\left\{
\prod_{j \leq a}
(z_j-q^{-2}w_a)
\prod_{a \leq j}
(w_a-q^4z_j)
\right\}}\nonumber\\
&\times&
\frac{\displaystyle
\prod_{1\leq a<b \leq |A|-\frac{M}{2}}
(v_a-v_b)(v_a-q^{-2}v_b)
\prod_{a,b \in A
\atop{a<b}}(w_a-w_b)(w_a-q^2w_b)
}{
\displaystyle
\prod_{b=1}^{|A|-\frac{M}{2}}
\prod_{a \in A}(v_b-qw_a)(v_b-q^{-1}w_a)}
\label{eqn:VEV-boson-1}
\\
&\times&
~_B\langle i|
e^{
\sum_{j=1}^{M}P(z_j)
+\sum_{a \in A}R^-(w_a)
+\sum_{b=1}^{|A|-\frac{M}{2}}R^+(v_b)}
e^{
\sum_{j=1}^{M}Q(z_j)
+\sum_{a \in A}S^-(w_a)
+\sum_{b=1}^{|A|-\frac{M}{2}}S^+(v_b)}
|i\rangle_B.\nonumber
\end{eqnarray}
Next we calculate the following vacuum expectation value more explicitly.
\begin{eqnarray}
&&
~_B\langle i|
e^{
\sum_{j=1}^{M}P(z_j)
+\sum_{a \in A}R^-(w_a)
+\sum_{b=1}^{|A|-\frac{M}{2}}R^+(v_b)}
e^{
\sum_{j=1}^{M}Q(z_j)
+\sum_{a \in A}S^-(w_a)
+\sum_{b=1}^{|A|-\frac{M}{2}}S^+(v_b)}
|i\rangle_B
\nonumber\\
&=&
\langle i|
e^{G^{(i)}}
e^{\sum_{n=1}^\infty a_{-n}X_n}
e^{-\sum_{n=1}^\infty a_n Y_n}
e^{F^{(i)}} |i \rangle.
\end{eqnarray}
Here we have used
\begin{eqnarray}
X_n&=&\frac{q^{{7 n}/{2}}}{[2 n]_q}\sum_{j=1}^{M}
z_j^{n}-\frac{q^{{n}/{2}}}{[n]_q}\sum_{a \in A}w_a^n+
\frac{q^{-{n}/{2}}}{[n]_q}\sum_{b=1}^{|A|-\frac{M}{2}} v_b^n,
\\
Y_n&=&
\frac{q^{-{5n}/{2}}}{[2n]_q}\sum_{j=1}^{M}
z_j^{-n}-
\frac{q^{{n}/{2}}}{[n]_q}
\sum_{a \in A}w_a^{-n}+\frac{q^{-{n}/{2}}}{[n]_q}
\sum_{b=1}^{|A|-\frac{M}{2}}
v_b^{-n}.
\end{eqnarray}
Using the relation $~_B\langle +;i|i;+ \rangle_B~=~_B\langle i|i\rangle_B$,
we have the following formula in \cite{JKKKM}:
\begin{eqnarray}
&&\frac{\langle i|e^{G^{(i)}}
e^{\sum_{n=1}^\infty a_{-n}X_n}e^{-\sum_{n=1}^\infty a_n Y_n}e^{F^{(i)}}|i \rangle}
{~_B\langle +;i|i;+ \rangle_B}\\
&=&
\exp\left(
\sum_{n=1}^\infty
\frac{[2n]_q[n]_q}{n}\frac{1}{1-\alpha_n \gamma_n}
\left\{\frac{1}{2}\gamma_n X_n^2-\alpha_n \gamma_n X_n Y_n +\frac{1}{2}\alpha_n Y_n^2+(\delta_n^{(i)}
+\gamma_n \beta_n^{(i)})X_n-(\beta_n^{(i)}+\alpha_n \delta_n^{(i)})Y_n
\right\}\right).\nonumber
\end{eqnarray}
Here $\alpha_n=-q^{6n}$, $\gamma_n=-q^{-2n}$, $\beta_n^{(i)}$, $\delta_n^{(i)}$ 
are given in (\ref{def:alpha}), (\ref{def:beta}), (\ref{def:delta}), respectively.
Note that the following infinite product relation
\begin{eqnarray}
\exp\left(-\sum_{n=1}^\infty \frac{1}{n}\frac{z^n}{(1-p_1^{n})(1-p_2^{n})\cdots (1-p^n)}\right)=
(z;p_1,p_2,\cdots,p_N)_\infty
\end{eqnarray}
has been used, where we denote
\begin{eqnarray}
(z;p_1,p_2,\cdots,p_N)_\infty=\prod_{n_1,n_2,\cdots,n_N=0}^\infty
(1-p_1^{n_1}p_2^{n_2}\cdots p_N^{n_N}z).
\end{eqnarray}
Below, we introduce the double-infinite products
\begin{eqnarray}
\{z\}_\infty=(z;q^4,q^4)_\infty,~~~[z]_\infty=(z;q^8,q^8)_\infty.
\label{def:infinite2}
\end{eqnarray}
We have following infinite product formula of the vacuum expectation value.
Note that the formulae summarized in Appendix \ref{appendix:C} are convenient for these calculations.
\begin{eqnarray}
&&
\frac{
\langle i|e^{G^{(i)}}e^{\sum_{n=1}^\infty a_{-n}X_n}
e^{-\sum_{n=1}^\infty a_n Y_n}e^{F^{(i)}}|i \rangle}{
~_B\langle +;i|i;+ \rangle_B}\nonumber\\
&=&
\left(\frac{\{q^6\}_\infty}{\{q^8\}_\infty}\right)^{M}
\{(q^4;q^2)_\infty\}^{|A|}
\{(q^2;q^2)_\infty\}^{|A|-\frac{M}{2}}
\nonumber\\
&\times&
\prod_{1\leq j<k \leq M}\frac{
\{q^6z_j z_k\}_\infty
\{q^2/z_jz_k\}_\infty
\{q^6z_j/z_k\}_\infty
\{q^6z_k/z_j\}_\infty
}{
\{q^8z_jz_k\}_\infty 
\{q^4/z_jz_k\}_\infty 
\{q^8z_j/z_k\}_\infty 
\{q^8z_k/z_j\}_\infty}\nonumber\\
&\times&
\prod_{j=1}^{M} \frac{
[q^{10}z_j^2]_\infty [q^{14}z_j^2]_\infty [q^{10}/z_j^2]_\infty [q^6 /z_j^2]_\infty}{
[q^{12}z_j^2]_\infty [q^{16}z_j^2]_\infty [q^{12}/z_j^2]_\infty [q^8/z_j^2]_\infty}\nonumber\\
&\times&
\frac{
\displaystyle
\prod_{j=1}^{M}\prod_{b=1}^{|A|-\frac{M}{2}} 
(qz_jv_b;q^4)_\infty (q^7z_j/v_b;q^4)_\infty 
(qv_b/z_j;q^4)_\infty (q^3/z_jv_b;q^4)_\infty
}{\displaystyle
\prod_{j=1}^{M} \prod_{a \in A}
(q^2 z_j w_a;q^4)_\infty 
(q^8 z_j/w_a;q^4)_\infty (q^2w_a/z_j;q^4)_\infty 
(q^4/z_jw_a;q^4)_\infty}\nonumber\\
&\times&
\frac{\displaystyle
\prod_{a \in A}(w_a^2/q^2;q^4)_\infty (q^6/w_a^2;q^4)_\infty
\prod_{b=1}^{|A|-\frac{M}{2}} (v_b^2/q^2;q^4)_\infty (q^6/v_b^2;q^4)_\infty}
{\displaystyle
\prod_{b=1}^{|A|-\frac{M}{2}} \prod_{a \in A}
(v_bw_a/q^3;q^2)_\infty 
(q^3 v_b/w_a;q^2)_\infty 
(q^3 w_a/v_b;q^2)_\infty 
(q^5/v_bw_a;q^2)_\infty
}
\nonumber\\
&\times&
\prod_{a,b \in A
\atop{a<b}}
(w_aw_b/q^2;q^2)_\infty 
(q^4w_a/w_b;q^2)_\infty (q^4w_b/w_a;q^2)_\infty 
(q^6/w_aw_b;q^2)_\infty\nonumber\\
&\times&
\prod_{1\leq a<b \leq {|A|-\frac{M}{2}}}
(v_av_b/q^4;q^2)_\infty 
(q^2v_a/v_b;q^2)_\infty 
(q^2v_b/v_a;q^2)_\infty 
(q^4/v_av_b;q^2)_\infty
\nonumber\\
&\times&
\left\{\begin{array}{cc}
\displaystyle
\prod_{j=1}^{M}\frac{(q^2rz_j;q^4)_\infty}{(q^4rz_j;q^4)_\infty}
\frac{\displaystyle \prod_{b=1}^{|A|-\frac{M}{2}} (1-rv_b/q^3)}{\displaystyle \prod_{a \in A}(1-q^{-2}rw_a)}& (i=0),\\
\displaystyle
\prod_{j=1}^{M}\frac{(1/rz_j;q^4)_\infty}{(q^2/rz_j;q^4)_\infty}
\frac{\displaystyle 
\prod_{b=1}^{|A|-\frac{M}{2}} (1-q/rv_b)}{\displaystyle
\prod_{a \in A}(1-q^2/rw_a)}& (i=1).
\end{array}\right.
\end{eqnarray}
Summarizing the above calculations, we finally obtain the following first integral representation
 of the $M$-point functions with $M$ even:
\begin{eqnarray}
&&
P_{\epsilon_1,\cdots,\epsilon_M}^{(+,i)}(\zeta_1,\cdots,\zeta_M;r,s)
=
P_{-\epsilon_1,\cdots,-\epsilon_M}^{(-,1-i)}(\zeta_1,\cdots,\zeta_M;1/r,-s/r)
\nonumber\\
&=&
(q s)^{|A|-\frac{M}{2}}(-q^3)^{\frac{1}{4}M^2-\sum_{a \in A}a}
\left(\frac{\{q^6\}_\infty}{\{q^8\}_\infty}\right)^{M}
(q^2;q^2)_\infty^{2|A|-\frac{M}{2}}
\nonumber\\
&\times&
\prod_{1\leq j<k \leq M}\frac{
\{q^6z_j z_k\}_\infty
\{q^2/z_jz_k\}_\infty
\{q^6z_j/z_k\}_\infty
\{q^2z_k/z_j\}_\infty
}{
\{q^8z_jz_k\}_\infty 
\{q^4/z_jz_k\}_\infty 
\{q^8z_j/z_k\}_\infty 
\{q^4z_k/z_j\}_\infty}
\prod_{j=1}^{M} 
\frac{[q^{10}z_j^2]_\infty [q^{14}z_j^2]_\infty [q^{10}/z_j^2]_\infty [q^6 /z_j^2]_\infty}{
[q^{12}z_j^2]_\infty [q^{16}z_j^2]_\infty [q^{12}/z_j^2]_\infty [q^8/z_j^2]_\infty}\nonumber\\
&\times&
\prod_{j=1}^M
\zeta_j^{\frac{1+\epsilon_j}{2}+M-j}
\sum_{l,m \geq 0 \atop{l+m=|A|-\frac{M}{2}}}
\frac{(-1)^l q^{-\frac{1}{2}l(l+1)+\frac{1}{2}m(m+1)}}
{[l]_q! [m]_q!}
\nonumber\\
&\times&
\oint \cdots \oint_{C_l^{(+,i)}} 
\prod_{a \in A} \frac{dw_a}{2\pi \sqrt{-1}}w_a^{1-i}
\prod_{b=1}^{|A|-\frac{M}{2}} \frac{dv_b}{2\pi \sqrt{-1}}v_b^{2i-1}
\frac{\displaystyle
\prod_{b=1}^{|A|-\frac{M}{2}}\prod_{j=1}^{M}
(v_b-q^3 z_j)
}{
\displaystyle
\prod_{a \in A}
\left\{
\prod_{1 \leq j \leq a}(z_j-q^{-2}w_a)
\prod_{a \leq j \leq M}(w_a-q^4z_j)
\right\}}\nonumber\\
&\times&
\frac{
\displaystyle
\prod_{j=1}^{M}\prod_{b=1}^{|A|-\frac{M}{2}} 
(qz_jv_b;q^4)_\infty (q^7z_j/v_b;q^4)_\infty 
(qv_b/z_j;q^4)_\infty (q^3/z_jv_b;q^4)_\infty
}{\displaystyle
\prod_{j=1}^{M} \prod_{a \in A}
(q^2 z_j w_a;q^4)_\infty 
(q^8 z_j/w_a;q^4)_\infty (q^2w_a/z_j;q^4)_\infty 
(q^4/z_jw_a;q^4)_\infty}\nonumber\\
&\times&
\frac{\displaystyle
\prod_{a \in A}(w_a^2/q^2;q^4)_\infty (q^6/w_a^2;q^4)_\infty
\prod_{b=1}^{|A|-\frac{M}{2}} (v_b^2/q^2;q^4)_\infty (q^6/v_b^2;q^4)_\infty}
{\displaystyle
\prod_{b=1}^{|A|-\frac{M}{2}} \prod_{a \in A}
v_b^2(v_bw_a/q^3;q^2)_\infty 
(q^3v_b/w_a;q^2)_\infty 
(w_a/q v_b;q^2)_\infty 
(q^5/v_b w_a;q^2)_\infty
}
\nonumber\\
&\times&
\prod_{a,b \in A
\atop{a<b}}
w_a^2 (w_aw_b/q^2;q^2)_\infty 
(q^4w_a/w_b;q^2)_\infty (w_b/w_a;q^2)_\infty 
(q^6/w_aw_b;q^2)_\infty\nonumber\\
&\times&
\prod_{1\leq a<b \leq |A|-\frac{M}{2}}
v_a^2 
(v_av_b/q^4;q^2)_\infty 
(q^2v_a/v_b;q^2)_\infty 
(q^{-2}v_b/v_a;q^2)_\infty 
(q^4/v_av_b;q^2)_\infty
\nonumber\\
&\times&
\left\{\begin{array}{cc}
\displaystyle
\prod_{j=1}^{M}\frac{(q^2rz_j;q^4)_\infty}{(q^4rz_j;q^4)_\infty}
\frac{\displaystyle \prod_{b=1}^{|A|-\frac{M}{2}} (1-rv_b/q^3)}{\displaystyle \prod_{a \in A}(1-q^{-2}rw_a)}& (i=0),\\
(-q^2 r)^{\frac{M}{2}-|A|}(-q^3)^\frac{M}{2} \displaystyle
\prod_{j=1}^{M}\zeta_j
\frac{(1/rz_j;q^4)_\infty}{(q^2/rz_j;q^4)_\infty}
\frac{\displaystyle 
\prod_{b=1}^{|A|-\frac{M}{2}} (1-q/rv_b)}{\displaystyle
\prod_{a \in A}(1-q^2/rw_a)}& (i=1).
\end{array}\right.
\label{eqn:correlation1}
\end{eqnarray}
Here we have used $\{z\}_\infty$ and $[z]_\infty$ defined in (\ref{def:infinite2}).
Recall that the set $A$ is given in (\ref{def:A}).
Here the integration contour $C_{l}^{(+,0)}$
is a simple closed curve 
that satisfies the following conditions for $s=0,1,2,\cdots$.
We set $L=|A|-\frac{M}{2}$.
The $w_a~(a \in A)$ encircles 
$q^{8+4s}z_j~(1\leq j < a)$, $q^{4+4s}z_j~(a \leq j \leq M)$, $q^{4+4s}/z_j~(1\leq j \leq M)$,
$q^{3+2s}v_b~(1\leq b \leq l)$, $q^{-1+2s}v_b~(l< b \leq L)$, $q^{5+2s}/v_b~(1\leq b \leq L)$,
but not
$q^{2-4s}z_j~(1\leq j \leq a)$, $q^{-2-4s}z_j~(a< j \leq M)$, $q^{-2-4s}/z_j~(1\leq j \leq M)$,
$q^{1-2s}v_b~(1\leq b \leq l)$, $q^{-3-2s}v_b~(l< b \leq L)$, $q^{3-2s}/v_b~(1\leq b \leq L)$, $q^2/r$.
The $v_b~(1\leq b \leq l)$ encircles 
$q^{-1+2s}w_a$, $q^{5+2s}/w_a~(a \in A)$ but not
$q^{-3-2s}w_a$, $q^{3-2s}/w_a~(a \in A)$.
The $v_b~(l< b \leq L)$
encircles
$q^{3+2s}w_a$, $q^{5+2s}/w_a~(a \in A)$ but not
$q^{1-2s}w_a$, $q^{3-2s}/w_a~(a \in A)$.
Similarly, the integration contour $C_l^{(+,1)}$
is a simple closed curve such that
$w_a~(a \in A)$ encircles $q^2/r$ in addition the same points as $C_l^{(+,0)}$ does.

\subsection{Second integral representation}

In this Section we consider $M$-point functions with $M$ an even integer :
\begin{eqnarray}
P^{(-,i)}_{\epsilon_1, \cdots, \epsilon_M}(\zeta_1,\cdots,\zeta_M;r,s)=
P^{(+,1-i)}_{-\epsilon_1, \cdots ,-\epsilon_M}(\zeta_1,\cdots,\zeta_M;1/r,-s/r).
\end{eqnarray}
Our aim is to calculate the vacuum expectation values using bosonizations 
associated with the lower-triangular boundary model $H_B^{(-)}$.
Recall that $A=\{1\leq a \leq M|\epsilon_a=+\}$.
From the bosonizations of the Drinfeld realization in Appendix \ref{appendix:B}, we have
\begin{eqnarray}
e_0 q^{-h_0}=q^{-1}x_1^-=q^{-1} \oint \frac{dv}{2\pi \sqrt{-1}}v X^-(v),~~~
f_1=x_0^-=\oint \frac{dv}{2\pi \sqrt{-1}}X^-(v).
\end{eqnarray}
From the normal orderings in appendix \ref{appendix:B}, we have
\begin{eqnarray}
(e_0 q^{-h_0})^l&=&q^{-l}
\prod_{b=1}^l \oint \frac{dv_b}{2\pi \sqrt{-1}} v_b
\prod_{1\leq a<b \leq l}
(v_a-v_b)(v_a-q^2 v_b)
:\prod_{b=1}^l X^-(v_b):,
\\
(f_1)^l&=&
\prod_{b=1}^l \oint \frac{dv_b}{2 \pi \sqrt{-1}}
\prod_{1\leq a<b \leq l}
(v_a-v_b)(v_a-q^2 v_b):\prod_{b=1}^lX^-(v_b):.
\end{eqnarray}
We have the following normal orderings
\begin{eqnarray}
&&
\exp_{q^{-1}}
\left(\frac{~s~}{q}e_0 q^{-h_0}\right)
\Phi_{\epsilon_1}^{(0,1)}(\zeta_1)\Phi_{\epsilon_2}^{(1,0)}(\zeta_2)\cdots
\Phi_{\epsilon_{M}}^{(1,0)}(\zeta_{M})
\exp_{q}\left(-\frac{~s~}{q}e_0 q^{-h_0} \right)\nonumber\\
&=&
(-q^3)^{\frac{1}{4}M(M-1)-\sum_{a \in A} a}(1-q^2)^{|A|}
\prod_{j=1}^{M}
\zeta_j^{\frac{1+\epsilon_j}{2}+M-j}
\prod_{1\leq j<k\leq M}\frac{(q^2z_k/z_j;q^4)_\infty}{
(q^4z_k/z_j;q^4)_\infty}
\nonumber\\
&\times&
\sum_{n=0}^\infty (s/q^2)^n
\sum_{l,m \geq 0
\atop{l+m=n}}\frac{(-1)^{m}
q^{-\frac{1}{2}l(l-1)+\frac{1}{2}m(m-1)}(-q^3)^{-M m}}{[l]_q! [m]_q!}\times
\prod_{b=1}^n
\oint \frac{dv_b}{2\pi \sqrt{-1}}v_b
\prod_{a \in A}\oint \frac{dw_a}{2\pi \sqrt{-1}}w_a\nonumber\\
&\times&
\frac{\displaystyle
\prod_{a,b \in A
\atop{a<b}}
(w_a-w_b)(w_a-q^2 w_b)
}{\displaystyle
\prod_{a \in A}
\left\{
\prod_{j \leq a}(z_j-q^{-2}w_a)
\prod_{a \leq j}(w_a-q^4 z_j)
\right\}}
\frac
{\displaystyle
\prod_{1\leq a<b\leq n}(v_a-v_b)(v_a-q^2v_b)}
{\displaystyle
\prod_{j=1}^{M}
\left\{
\prod_{b=1}^l(v_b-q^4z_j)
\prod_{b=l+1}^n(z_j-q^{-2}v_b)
\right\}}\nonumber\\
&\times&
\prod_{b=1}^l\prod_{a \in A}
(v_b-w_a)(v_b-q^2 w_a)
\prod_{a \in A}
\prod_{b=l+1}^n
(w_a-v_b)
(w_a-q^2 v_b)\nonumber\\
&\times&
:\Phi_{-}^{(0,1)}(\zeta_1)\Phi_{-}^{(1,0)}(\zeta_2)\cdots
\Phi_{-}^{(1,0)}(\zeta_{M})
\prod_{a \in A}X^-(w_a)
\prod_{b=1}^n X^-(v_b):,
\end{eqnarray}
and
\begin{eqnarray}
&&\exp_{q}
\left(\frac{~s~}{r}f_1 \right)
\Phi_{\epsilon_1}^{(1,0)}(\zeta_1)\Phi_{\epsilon_2}^{(0,1)}(\zeta_2)\cdots
\Phi_{\epsilon_{M}}^{(0,1)}(\zeta_{M})
\exp_{q^{-1}}\left(-\frac{~s~}{r} f_1 \right)\nonumber\\
&=&
(-q^3)^{\frac{1}{4}M(M-1)-\sum_{a \in A} a}(1-q^2)^{|A|}
\prod_{j=1}^{M}
\zeta_j^{\frac{1+\epsilon_j}{2}+M-j}
\prod_{1\leq j<k\leq M}\frac{(q^2z_k/z_j;q^4)_\infty}{
(q^4z_k/z_j;q^4)_\infty}
\nonumber\\
&\times&
\sum_{n=0}^\infty (s/r)^n
\sum_{l,m \geq 0
\atop{l+m=n}}\frac{(-1)^{m}
q^{\frac{1}{2}l(l-1)-\frac{1}{2}m(m-1)}(-q^3)^{-M m}}{[l]_q! [m]_q!}\times
\prod_{b=1}^n
\oint \frac{dv_b}{2\pi \sqrt{-1}}
\prod_{a \in A}\oint \frac{dw_a}{2\pi \sqrt{-1}}w_a\nonumber\\
&\times&
\frac{\displaystyle
\prod_{a,b \in A
\atop{a<b}}
(w_a-w_b)(w_a-q^2 w_b)
}{\displaystyle
\prod_{a \in A}
\left\{
\prod_{j \leq a}(z_j-q^{-2}w_a)
\prod_{a \leq j}(w_a-q^4 z_j)
\right\}}
\frac
{\displaystyle
\prod_{1\leq a<b\leq n}(v_a-v_b)(v_a-q^2v_b)}
{\displaystyle
\prod_{j=1}^{M}
\left\{
\prod_{b=1}^l(v_b-q^4z_j)
\prod_{b=l+1}^n(z_j-q^{-2}v_b)
\right\}}\nonumber\\
&\times&
\prod_{a=1}^l\prod_{b \in A}
(v_a-w_b)(v_a-q^2 w_b)
\prod_{a \in A}
\prod_{b=l+1}^n
(w_a-v_b)
(w_a-q^2 v_b)\nonumber\\
&\times&
:\Phi_{-}^{(0,1)}(\zeta_1)\Phi_{-}^{(1,0)}(\zeta_2)\cdots
\Phi_{-}^{(1,0)}(\zeta_{M})
\prod_{a \in A}X^-(w_a)
\prod_{b=1}^n X^-(v_b):.
\end{eqnarray}
The zero-mode $e^{\alpha}$ part of the operator
$:\Phi_-^{(i,1-i)}(\zeta_1)\cdots \Phi_-^{(1-i,i)}(\zeta_M)
\prod_{a \in A}X^-(w_a)\prod_{b=1}^n X^-(v_b):$
is given by $e^{\alpha(\frac{M}{2}-|A|-n)}$.
Hence, the condition for which the vacuum expectation value  is non-vanishing reads
\begin{eqnarray}
&&~_B\langle i|:\Phi_{-}^{(i,1-i)}(\zeta_1)\cdots
\Phi_{-}^{(1-i,i)}(\zeta_{M})
\prod_{a \in A}X^-(w_a)
\prod_{b=1}^n X^-(v_b):|i\rangle_B \neq 0,
\nonumber\\
&&
\Longleftrightarrow 
~~_B\langle i|e^{(\frac{M}{2}-|A|-n)\alpha}|i\rangle_B \neq 0~~
\Longleftrightarrow~~~
\frac{M}{2}-|A|=n\geq 0.
\label{prop:operator-zero-2}
\end{eqnarray}
For $\frac{M}{2} \geq |A|$, it implies that non-zero vacuum expectation value takes the form
\begin{eqnarray}
&&~_B\langle -;i|
\Phi_{\epsilon_1}^{(i,1-i)}(\zeta_1)\Phi_{\epsilon_2}^{(1-i,i)}(\zeta_2)
\cdots
\Phi_{\epsilon_{M}}^{(i,1-i)}(\zeta_{M})
|i;-\rangle_B
\nonumber\\
&=&
(-q^3)^{\frac{1}{4}M^2+\frac{i}{2}M-\sum_{a \in A} a}(1-q^2)^{|A|}
(s/q^2)^{\frac{1}{2}M-|A|}
(q^2/r)^{(\frac{M}{2}-|A|)i}
\prod_{j=1}^{M}
\zeta_j^{\frac{1+\epsilon_j}{2}+M-j+i}
\prod_{1\leq j<k\leq M}\frac{(q^2z_k/z_j;q^4)_\infty}{
(q^4z_k/z_j;q^4)_\infty}
\nonumber\\
&\times&
\sum_{l,m \geq 0
\atop{l+m=\frac{M}{2}-|A|}}\frac{(-1)^{m(M+1)}
q^{\{-\frac{1}{2}l(l-1)+\frac{1}{2}m(m-1)\}(1-2 i)-3 M m}}{[l]_q! [m]_q!}\times
\prod_{b=1}^{\frac{M}{2}-|A|}
\oint \frac{dv_b}{2\pi \sqrt{-1}}v_b^{1-2i}
\prod_{a \in A}\oint \frac{dw_a}{2\pi \sqrt{-1}}
w_a^{1-i}\nonumber\\
&\times&
\frac{\displaystyle
\prod_{a,b \in A
\atop{a<b}}
(w_a-w_b)(w_a-q^2 w_b)
}{\displaystyle
\prod_{a \in A}
\left\{
\prod_{1 \leq j \leq a}(z_j-q^{-2}w_a)
\prod_{a \leq j \leq M}(w_a-q^4 z_j)
\right\}}
\frac
{\displaystyle
\prod_{1\leq a<b\leq \frac{M}{2}-|A|}(v_a-v_b)(v_a-q^2v_b)}
{\displaystyle
\prod_{j=1}^{M}
\left\{
\prod_{b=1}^l(v_b-q^4z_j)
\prod_{b=l+1}^{\frac{M}{2}-|A|}(z_j-q^{-2}v_b)
\right\}}\nonumber\\
&\times&
\prod_{b=1}^l\prod_{a \in A}
(v_b-w_a)(v_b-q^2 w_a)
\prod_{a \in A}
\prod_{b=l+1}^{\frac{M}{2}-|A|}
(w_a-v_b)
(w_a-q^2 v_b)\nonumber\\
&\times&
~_B\langle i|
e^{
\sum_{j=1}^{M}P(z_j)
+\sum_{a \in A}R^-(w_a)
+\sum_{b=1}^{\frac{M}{2}-|A|}R^-(v_b)}
e^{
\sum_{j=1}^{M}Q(z_j)
+\sum_{a \in A}S^-(w_a)
+\sum_{b=1}^{\frac{M}{2}-|A|}S^-(v_b)}
|i\rangle_B.
\label{eqn:VEV-boson-2}
\end{eqnarray}
Next, we calculate the vacuum expectation value more explicitly.
\begin{eqnarray}
&&
~_B\langle i|
e^{
\sum_{j=1}^{M}P(z_j)
+\sum_{a \in A}R^-(w_a)
+\sum_{a=1}^{\frac{M}{2}-|A|}R^-(v_a)}
e^{
\sum_{j=1}^{M}Q(z_j)
+\sum_{a \in A}S^-(w_a)
+\sum_{a=1}^{\frac{M}{2}-|A|}S^-(v_a)}
|i\rangle_B
\nonumber\\
&=&
\langle i|
e^{G^{(i)}}
e^{\sum_{n=1}^\infty a_{-n}X_n}
e^{-\sum_{n=1}^\infty a_n Y_n}
e^{F^{(i)}} |i \rangle.
\end{eqnarray}
Here we have defined
\begin{eqnarray}
X_n&=&\frac{q^{{7 n}/{2}}}{[2 n]_q}\sum_{j=1}^{M}
z_j^{n}-\frac{q^{{n}/{2}}}{[n]_q}\sum_{a \in A}w_a^n-
\frac{q^{{n}/{2}}}{[n]_q}
\sum_{b=1}^{\frac{M}{2}-|A|} v_b^n,
\\
Y_n&=&
\frac{q^{-{5n}/{2}}}{[2n]_q}\sum_{j=1}^{M}
z_j^{-n}-
\frac{q^{{n}/{2}}}{[n]_q}
\sum_{a \in A}w_a^{-n}-\frac{q^{{n}/{2}}}{[n]_q}
\sum_{b=1}^{\frac{M}{2}-|A|}
v_b^{-n}.
\end{eqnarray}
By straightforward calculations, the  following infinite product formula of the vacuum expectation value is obtained:
\begin{eqnarray}
&&\frac{
\langle i|e^{G^{(i)}}
e^{\sum_{n=1}^\infty a_{-n}X_n}
e^{-\sum_{n=1}^\infty a_n Y_n}e^{F^{(i)}}|i \rangle}{
~_B\langle +;i|i;+ \rangle_B}=
\left(\frac{\{q^6\}_\infty}{\{q^8\}_\infty}\right)^{M}
\{(q^4;q^2)_\infty\}^{\frac{M}{2}}
\nonumber\\
&\times&
\prod_{1\leq j<k \leq M}\frac{
\{q^6z_j z_k\}_\infty
\{q^2/z_jz_k\}_\infty
\{q^6z_j/z_k\}_\infty
\{q^6z_k/z_j\}_\infty
}{
\{q^8z_jz_k\}_\infty 
\{q^4/z_jz_k\}_\infty 
\{q^8z_j/z_k\}_\infty 
\{q^8z_k/z_j\}_\infty}
\prod_{j=1}^{M} \frac{
[q^{10}z_j^2]_\infty [q^{14}z_j^2]_\infty [q^{10}/z_j^2]_\infty [q^6 /z_j^2]_\infty}{
[q^{12}z_j^2]_\infty [q^{16}z_j^2]_\infty [q^{12}/z_j^2]_\infty [q^8/z_j^2]_\infty}\nonumber\\
&\times&
\frac{
\displaystyle
\prod_{a \in A}(w_a^2/q^2;q^4)_\infty (q^6/w_a^2;q^4)_\infty
}{\displaystyle
\prod_{j=1}^{M}
\prod_{a \in A}
(q^2 z_j w_a;q^4)_\infty 
(q^8 z_j/w_a;q^4)_\infty (q^2w_a/z_j;q^4)_\infty 
(q^4/z_jw_a;q^4)_\infty}\nonumber\\
&\times&\frac{
\displaystyle
\prod_{b=1}^{\frac{M}{2}-|A|} 
(v_b^2/q^2;q^4)_\infty (q^6/v_b^2;q^4)_\infty
}{\displaystyle
\prod_{j=1}^{M}
\prod_{b=1}^{\frac{M}{2}-|A|}
(q^2 z_j v_b;q^4)_\infty 
(q^8 z_j/v_b;q^4)_\infty (q^2v_b/z_j;q^4)_\infty 
(q^4/z_jv_b;q^4)_\infty
}\nonumber\\
&\times&
\prod_{b=1}^{\frac{M}{2}-|A|} \prod_{a \in A}
(v_bw_a/q^2;q^2)_\infty 
(q^4 v_b/w_a;q^2)_\infty 
(q^4 w_a/v_b;q^2)_\infty 
(q^6/v_bw_a;q^2)_\infty
\nonumber\\
&\times&
\prod_{a,b \in A
\atop{a<b}}
(w_aw_b/q^2;q^2)_\infty 
(q^4w_a/w_b;q^2)_\infty (q^4w_b/w_a;q^2)_\infty 
(q^6/w_aw_b;q^2)_\infty\nonumber\\
&\times&
\prod_{1\leq a<b \leq L}
(v_av_b/q^2;q^2)_\infty 
(q^4v_a/v_b;q^2)_\infty 
(q^4v_b/v_a;q^2)_\infty 
(q^6/v_av_b;q^2)_\infty
\nonumber\\
&\times&
\left\{\begin{array}{cc}
\displaystyle
\prod_{j=1}^{M}\frac{(q^2rz_j;q^4)_\infty}{(q^4rz_j;q^4)_\infty}
\frac{1}{\displaystyle 
\prod_{a \in A}(1-rw_a/q^2)
\prod_{b=1}^{\frac{M}{2}-|A|} (1-rv_b/q^2)
}& (i=0),\\
\displaystyle
\prod_{j=1}^{M}\frac{(1/rz_j;q^4)_\infty}{(q^2/rz_j;q^4)_\infty}
\frac{1}{\displaystyle
\prod_{a \in A}(1-q^2/rw_a)
\prod_{b=1}^{\frac{M}{2}-|A|} (1-q^2/rv_b)
}& (i=1).
\end{array}\right.
\end{eqnarray}
Summarizing the above calculations, we finally obtain the  second
integral representation of the $M$-point functions with $M$ even.
\begin{eqnarray}
&&
P_{\epsilon_1,\cdots,\epsilon_M}^{(-,i)}(\zeta_1,\cdots,\zeta_m;r,s)=
P_{-\epsilon_1,\cdots,-\epsilon_M}^{(+,1-i)}(\zeta_1,\cdots,\zeta_M;1/r,-s/r)
\nonumber\\
&=&
(-q^3)^{\frac{1}{4}M^2-\sum_{a \in A}a}(s/q^2)^{\frac{M}{2}-|A|}
(1-q^2)^{|A|}\left(\frac{\{q^6\}_\infty}{\{q^8\}_\infty}\right)^{M}
\{(q^4;q^2)_\infty\}^{\frac{M}{2}}
\nonumber\\
&\times&
\prod_{1\leq j<k \leq M}\frac{
\{q^6z_j z_k\}_\infty
\{q^2/z_jz_k\}_\infty
\{q^6z_j/z_k\}_\infty
\{q^2z_k/z_j\}_\infty
}{
\{q^8z_jz_k\}_\infty 
\{q^4/z_jz_k\}_\infty 
\{q^8z_j/z_k\}_\infty 
\{q^4z_k/z_j\}_\infty}
\prod_{j=1}^{M} 
\frac{
[q^{10}z_j^2]_\infty [q^{14}z_j^2]_\infty [q^{10}/z_j^2]_\infty [q^6 /z_j^2]_\infty}{
[q^{12}z_j^2]_\infty [q^{16}z_j^2]_\infty [q^{12}/z_j^2]_\infty [q^8/z_j^2]_\infty}\nonumber\\
&\times&
\prod_{j=1}^{M}\zeta_j^{\frac{1+\epsilon_j}{2}+M-j}
\sum_{l,m \geq 0
\atop{l+m=\frac{M}{2}-|A|}}
\frac{(-1)^{m(M+1)} q^{\{-\frac{1}{2}l(l-1)+\frac{1}{2}m(m-1)\}(1-2i)-3M m}}{[l]_q! [m]_q!}\nonumber\\
&\times&
\oint \cdots \oint_{C_l^{(-,i)}} 
\prod_{b=1}^{\frac{M}{2}-|A|}\frac{dv_b}{2\pi \sqrt{-1}}v_b^{1-2i}
\prod_{a \in A}\frac{dw_a}{2\pi \sqrt{-1}}w_a^{1-i}
\nonumber\\
&\times&
\frac{
\displaystyle
\prod_{a \in A}(w_a^2/q^2;q^4)_\infty (q^6/w_a^2;q^4)_\infty
\prod_{b=1}^{\frac{M}{2}-|A|} 
(v_b^2/q^2;q^4)_\infty (q^6/v_b^2;q^4)_\infty
}{\displaystyle
\prod_{a\in A}
\left\{\prod_{1 \leq j \leq a}(z_j-q^{-2}w_a)
\prod_{a \leq j \leq M}(w_a-q^4z_j)
\right\}
\prod_{j=1}^{M}
\left\{
\prod_{b=1}^{l}
(v_b-q^4z_j)
\prod_{b=l+1}^{\frac{M}{2}-|A|}(z_j-q^{-2}v_b)\right\}
}\nonumber\\
&\times&
\frac{
\displaystyle
\prod_{a,b \in A
\atop{a<b}}w_a^2 
(w_aw_b/q^2;q^2)_\infty 
(q^4w_a/w_b;q^2)_\infty 
(w_b/w_a;q^2)_\infty 
(q^6/w_aw_b;q^2)_\infty
}{\displaystyle
\prod_{j=1}^{M}
\prod_{a \in A}
(q^2 z_j w_a;q^4)_\infty 
(q^8 z_j/w_a;q^4)_\infty (q^2w_a/z_j;q^4)_\infty 
(q^4/z_jw_a;q^4)_\infty}\nonumber\\
&\times&\frac{
\displaystyle
\prod_{1\leq a<b \leq \frac{M}{2}-|A|}
v_a^2 (v_av_b/q^2;q^2)_\infty 
(q^4v_a/v_b;q^2)_\infty 
(v_b/v_a;q^2)_\infty 
(q^6/v_av_b;q^2)_\infty
}{\displaystyle
\prod_{j=1}^{M}
\prod_{b=1}^{\frac{M}{2}-|A|}
(q^2 z_j v_b;q^4)_\infty 
(q^8 z_j/v_b;q^4)_\infty (q^2v_b/z_j;q^4)_\infty 
(q^4/z_jv_b;q^4)_\infty
}\nonumber\\
&\times&
\prod_{b=1}^{l} \prod_{a \in A}
v_b^2 (v_bw_a/q^2;q^2)_\infty 
(q^4 v_b/w_a;q^2)_\infty 
(w_a/v_b;q^2)_\infty 
(q^6/v_bw_a;q^2)_\infty
\nonumber\\
&\times&
\prod_{b=l+1}^{\frac{M}{2}-|A|} \prod_{a \in A}
w_a^2 (v_bw_a/q^2;q^2)_\infty 
(v_b/w_a;q^2)_\infty 
(q^4 w_a/v_b;q^2)_\infty 
(q^6/v_bw_a;q^2)_\infty
\nonumber\\
&\times&
\left\{\begin{array}{cc}
\displaystyle
\prod_{j=1}^{M}\frac{(q^2rz_j;q^4)_\infty}{(q^4rz_j;q^4)_\infty}
\frac{1}{\displaystyle 
\prod_{a \in A}(1-rw_a/q^2)
\prod_{b=1}^{\frac{M}{2}-|A|} (1-rv_b/q^2)
}& (i=0),\\
\displaystyle
(-q^3)^{\frac{M}{2}} (q^2/r)^{\frac{M}{2}-|A|}
\prod_{j=1}^{M}\zeta_j \frac{(1/rz_j;q^4)_\infty}{(q^2/rz_j;q^4)_\infty}
\frac{1}{\displaystyle
\prod_{a \in A}(1-q^2/rw_a)
\prod_{b=1}^{\frac{M}{2}-|A|} (1-q^2/rv_b)
}& (i=1).
\end{array}\right.
\label{eqn:correlation2}
\end{eqnarray}
Here we have used $\{z\}_\infty$ and $[z]_\infty$ defined in (\ref{def:infinite2}).  
Recall that the set $A$ is given in (\ref{def:A}).
Here the integration contour $C_l^{(-,0)}$ is a simple closed curve 
that satisfies the following conditions for $s=0,1,2,\cdots$.
The $w_a~(a \in A)$ encircles
$q^{8+4s}z_j~(1\leq j < a)$, $q^{4+4s}z_j~(a\leq j \leq M)$, $q^{4+4s}/z_j~(1\leq j \leq M)$,
but not
$q^{2-4s}z_j~(1\leq j \leq a)$, $q^{-2-4s}z_j~(a < j \leq M)$, $q^{-2-4s}/z_j~(1\leq j \leq M)$, $q^2/r$.
The $v_b~(1\leq b \leq l)$ encircles
$q^{4+4s}z_j,~q^{4+4s}/z_j~(1\leq j \leq M)$,
but~not
$q^{-2-4s}z_j,~q^{-2-4s}/z_j~(1\leq j \leq M)$, $q^2/r$.
The $v_b~(l<b\leq \frac{M}{2}-|A|)$
encircles 
$q^{8+4s}z_j, q^{4+4s}/z_j~(1\leq j \leq M)$ but not
$q^{2-4s}z_j, q^{-2-4s}/z_j~(1\leq j \leq M)$, $q^2/r$.
The integration contour $C_l^{(-,1)}$
is a simple closed curve such that
$w_a~(a \in A)$ encircles $q^2/r$ 
and $v_b~(1\leq b \leq \frac{M}{2}-|A|)$
encircles $q^2/r$ in addition the same points as $C_l^{(+,0)}$ does.

\subsection{Diagonal degeneration}

\label{section:correlation-diagonal}

The purpose of this subsection is to identify a sufficient condition such that
the expressions for a triangular boundary condition coincide with those associated with a  diagonal boundary condition.
Let us go back to the formula (\ref{eqn:VEV-boson-1}).
We note that $\sum_{j=1}^M \epsilon_j=0 \Leftrightarrow n=|A|-\frac{M}{2}=0$.
Upon the specialization 
$n=|A|-\frac{M}{2}=0$, we have
\begin{eqnarray}
~_B\langle i;\pm|\Phi_{\epsilon_1}^{(i,1-i)}(\zeta_1)
\cdots
\Phi_{\epsilon_{M}}^{(1-i,i)}(\zeta_{M})
|\pm;i\rangle_B=
~_B\langle i|\Phi_{\epsilon_1}^{(i,1-i)}(\zeta_1)
\cdots
\Phi_{\epsilon_{M}}^{(1-i,i)}(\zeta_{M})
|i\rangle_B.
\end{eqnarray}
The same argument holds for (\ref{eqn:VEV-boson-2}).
We conclude that
upon the parity preserving condition
\begin{eqnarray}
\epsilon_1+\epsilon_2+\cdots +\epsilon_M=0,\label{eqn:sum-eps}
\end{eqnarray} 
we have the same integral representation as the one for the diagonal boundary conditions \cite{JKKKM}.
Here we note that we have revised misprints in (4.8) of \cite{JKKKM}.
\begin{eqnarray}
&&
\frac{~_B\langle i;\pm|\Phi_{\epsilon_1}^{(i,1-i)}(\zeta_1)
\cdots
\Phi_{\epsilon_{M}}^{(1-i,i)}(\zeta_{M})
|\pm;i\rangle_B
}{~_B\langle i;\pm|\pm;i\rangle_B}
\nonumber\\
&=&(-q^3)^{\frac{1}{4}M^2-\sum_{a \in A}a}
\left(\frac{\{q^6\}_\infty}{\{q^8\}_\infty}\right)^{M} (q^2;q^2)_\infty^{\frac{M}{2}}
\prod_{j=1}^{M} \zeta_j^{\frac{1+\epsilon_j}{2}-j+M}\nonumber\\
&\times&
\prod_{1\leq j<k \leq M}\frac{
\{q^6z_j z_k\}_\infty
\{q^2/z_jz_k\}_\infty
\{q^6z_j/z_k\}_\infty
\{q^2z_k/z_j\}_\infty
}{
\{q^8z_jz_k\}_\infty 
\{q^4/z_jz_k\}_\infty 
\{q^8z_j/z_k\}_\infty 
\{q^4z_k/z_j\}_\infty}
\prod_{j=1}^{M} \frac{
[q^{10}z_j^2]_\infty [q^{14}z_j^2]_\infty [q^{10}/z_j^2]_\infty [q^6 /z_j^2]_\infty}{
[q^{12}z_j^2]_\infty [q^{16}z_j^2]_\infty [q^{12}/z_j^2]_\infty [q^8/z_j^2]_\infty}\nonumber\\
&\times&
\prod_{a \in A}
\oint_{C_0^{(i)}} \frac{dw_a}{2\pi \sqrt{-1}}
\frac{\displaystyle
\prod_{a,b \in A \atop{a<b}}
w_a^2 (q^{-2}w_aw_b;q^2)_\infty
(q^4w_a/w_b;q^2)_\infty 
(w_b/w_a;q^2)_\infty 
(q^6/w_aw_b;q^2)_\infty
}{\displaystyle
\prod_{a \in A}\prod_{j=1}^{M} 
(q^2 z_j w_a;q^4)_\infty
(q^8 z_j/w_a;q^4)_\infty 
(q^2 w_a/z_j;q^4)_\infty 
(q^4/z_j w_a;q^4)_\infty}
\nonumber
\\
&\times&
\frac{\displaystyle
\prod_{a \in A}
(q^{-2}w_a^2;q^4)_\infty 
(q^6/w_a^2;q^4)_\infty}{
\displaystyle
\prod_{a \in A}
\left\{\prod_{1\leq j \leq a}
(z_j-w_a/q^2)
\prod_{a \leq j\leq M}
(w_a-q^4z_j)\right\}}\nonumber\\
&\times&
\left\{\begin{array}{cc}
\displaystyle
\prod_{j=1}^{M} \frac{(q^2rz_j;q^4)_\infty}{(q^4rz_j;q^4)_\infty}
\prod_{a \in A}\frac{w_a}{(1-rw_a/q^2)}& ({\rm for }~i=0),\\
\displaystyle
(-q^3)^{\frac{M}{2}}
\prod_{j=1}^{M} \zeta_j\frac{(1/rz_j;q^4)_\infty}{(q^2/rz_j;q^4)_\infty}
\prod_{a \in A}\frac{1}{(1-q^2/ rw_a)}
&~({\rm for}~i=1).
\\
\end{array}\right.
\end{eqnarray}
Here we have used $\{z\}_\infty$ and $[z]_\infty$ defined in (\ref{def:infinite2}).  
The integration contour $C_0^{(0)}$ is a simple closed curve 
that satisfies the following conditions for $s=0,1,2,\cdots$.
The $w_a~(a \in A)$ encircles
$q^{8+4s}z_j~(1\leq j < a)$, $q^{4+4s}z_j~(a\leq j \leq M)$, $q^{4+4s}/z_j~(1\leq j \leq M)$,
but not
$q^{2-4s}z_j~(1\leq j \leq a)$, $q^{-2-4s}z_j~(a < j \leq M)$, $q^{-2-4s}/z_j~(1\leq j \leq M)$, $q^2/r$.
The integration contour $C_0^{(1)}$
is a simple closed curve such that
$w_a~(a \in A)$ encircles $q^2/r$ 
in addition the same points as $C_0^{(0)}$ does.

To conclude, let us mention that according to the zero-mode operators 
(\ref{prop:operator-zero-1}) and (\ref{prop:operator-zero-2}), we have
\begin{eqnarray}
P_{\epsilon_1,\cdots,\epsilon_M}^{(\pm,i)}(\zeta_1,\cdots,\zeta_M;r,s)=0~~~{\rm for}~~\pm\left(\frac{M}{2}-|A|\right)>0.
\end{eqnarray}

\subsection{Boundary expectation values of spin operators}

In this Section we study simple examples.
First, let us consider the boundary expectation value given by
\begin{eqnarray}
\frac{~_B\langle i;-|\sigma^+_1|-;i\rangle_B}{
~_B\langle i;-|-;i\rangle_B}&=&
g \left.P_{-,-}^{(-,i)}(-q^{-1}\zeta,\zeta;r,s)\right|_{\zeta=1},
\end{eqnarray}
where, according to (\ref{eqn:correlation2}), we have:
\begin{eqnarray}
P_{-,-}^{(-,i)}(-q^{-1}\zeta,\zeta;r,s)
&=& (-s) \zeta
\frac{\Theta_{q^4}(\zeta^4)}{(1-\zeta^4)}
(q^2;q^2)_\infty^3
\oint_{C^{(-,i)}}
\frac{dv}{2\pi \sqrt{-1}}
v(1+v/q^2\zeta^2)(1-q^2/v \zeta^2)\nonumber\\
&\times&
\frac{\Theta_{q^4}(q^2v^2)}{\Theta_{q^2}(v\zeta^2)\Theta_{q^2}(v/\zeta^2)}
\left\{\begin{array}{cc}
\frac{\displaystyle(1-r\zeta^2)}{\displaystyle(1-rv/q^2)q^2}&(i=0),
\\
\frac{\displaystyle
(1-r\zeta^2)}{\displaystyle
(1-rv/q^2)r v}&(i=1).
\end{array}
\right.
\end{eqnarray}
Here the integration contour $C^{(-,0)}$ encircles $q^{2k}\zeta^2$, $q^{2k+2}/\zeta^2$ $(k=1,2,\cdots)$.
The integration contour $C^{(-,1)}$ encircles $q^{2k}\zeta^2$, $q^{2k+2}/\zeta^2$ $(k=1,2,\cdots)$, and $q^2/r$.
Here we have used following simplification
\begin{eqnarray}
\frac{(q^4;q^4)_\infty}{1-z^2}\Theta_{q^4}(z^2)
&=&
\left(\frac{\{q^6\}_\infty}{\{q^8\}_\infty}\right)^{2}
\frac{\{q^6z_1 z_2\}_\infty
\{q^2/z_1z_2\}_\infty
\{q^6z_1/z_2\}_\infty
\{q^2z_2/z_1\}_\infty
}{
\{q^8z_1z_2\}_\infty 
\{q^4/z_1z_2\}_\infty 
\{q^8z_1/z_2\}_\infty 
\{q^4z_2/z_1\}_\infty}\nonumber\\
&\times&\left.
\prod_{j=1}^{2} 
\frac{
[q^{10}z_j^2]_\infty [q^{14}z_j^2]_\infty [q^{10}/z_j^2]_\infty [q^6 /z_j^2]_\infty}{
[q^{12}z_j^2]_\infty [q^{16}z_j^2]_\infty [q^{12}/z_j^2]_\infty [q^8/z_j^2]_\infty}
\right|_{z_1=q^{-2}z,z_2=z}.
\end{eqnarray}
Using properties of the theta function :
$\frac{\Theta_{q^4}(q^2v^2)}{\Theta_{q^2}(v\zeta^2)\Theta_{q^2}(v/\zeta^2)}=(-q^{-2})^k
\frac{\Theta_{q^4}(q^{-4k}q^2v^2)}{\Theta_{q^2}(q^{-2k}v\zeta^2)\Theta_{q^2}(q^{-2k}v/\zeta^2)}$ and
$\Theta_p(1/z)=-\frac{1}{z}\Theta_p(z)$, we calculate the residues.
We have
\begin{eqnarray}
g P_{-,-}^{(-,0)}(-q^{-1}\zeta, \zeta)=
s \zeta
\left(
2+\frac{1-rz}{z}
\sum_{k=1}^\infty
(-q^2)^{k}\frac{
(z-z^{-1})-r(1+q^{4k})+(z+z^{-1})q^{2k}
}{(1-q^{2k}rz)(1-q^{2k}r/z)}
\right),
\end{eqnarray}
and
\begin{eqnarray}
g P_{-,-}^{(-,1)}(-q^{-1}\zeta,\zeta)
&=&
\frac{s \zeta }{r}\left(
\frac{2}{z}+\frac{1-rz}{z}
\sum_{k=1}^\infty (-1)^{k}
\frac{(z-z^{-1})rq^{2k}+q^{2k}+q^{-2k}-(z+z^{-1})r}{
(1-rq^{2k}z)(1-rq^{2k}/z)}
\right)\nonumber\\
&+&
\frac{s \zeta q^2}{r^2}
\frac{(q^2;q^2)_\infty^4}{(q^4;q^4)_\infty^2}
\frac{(1+1/rz)(1-r/z)(1-rz)}{1-z^2}
\frac{\Theta_{q^4}(z^2)\Theta_{q^4}(q^6/r^2)}{\Theta_{q^2}(q^2z/r)\Theta_{q^2}(q^2/zr)}.
\end{eqnarray}
Upon the specialization $\zeta=1$, the boundary expectation value of the spin operator $\sigma_1^+$ in the model $H_B^{(-)}$ is obtained:
\begin{eqnarray}
\frac{~_B\langle 0;-|
\sigma_1^+|-;0\rangle_B}
{~_B\langle 0;-|-;0\rangle_B}
&=&
s \left(
2+(1-r)\sum_{k=1}^\infty
(-q^2)^{k}\frac{2q^{2k}
-r(1+q^{4k})}{(1-rq^{2k})^2}
\right),\\
\frac{~_B\langle 1;-|
\sigma_1^+|-;1\rangle_B}
{~_B\langle 1;-|-;1\rangle_B}
&=&\frac{s}{r}
\left(2+(1-r)
\sum_{k=1}^\infty (-1)^{k}
\frac{q^{2k}+q^{-2k}-2r}{
(1-rq^{2k})^2}\right)\nonumber\\
&+&
\frac{s q^2}{r^2}(1+1/r)(q^4;q^4)_\infty^2(q^2;q^2)_\infty^2
\frac{(q^6/r^2;q^4)_\infty
(r^2/q^2;q^4)_\infty}{
(q^2/r;q^2)_\infty^2 (q^2r;q^2)_\infty^2}.
\end{eqnarray}
Then, using the spin-reversal property (\ref{eqn:reversal-correlation})  the boundary expectation value of the spin operator $\sigma_1^-$ in  the model $H_B^{(+)}$ immediately follows:
\begin{eqnarray}
\frac{~_B\langle 1;+|
\sigma_1^-|+;1\rangle_B}
{~_B\langle 1;+|+;1 \rangle_B}
&=&
-\frac{s}{r} \left(
2+(1-1/r)\sum_{k=1}^\infty
(-q^2)^{k}\frac{2q^{2k}-(1+q^{4k})/r}{(1-q^{2k}/r)^2}
\right),\\
\frac{~_B\langle 0;+|
\sigma_1^-|+;0\rangle_B}
{~_B\langle 0;+|+;0 \rangle_B}
&=&-s
\left(2+(1-1/r)
\sum_{k=1}^\infty (-1)^{k}
\frac{q^{2k}+q^{-2k}-2/r}{
(1-q^{2k}/r)^2}\right)\nonumber\\
&-&s q^2 r (1+r)(q^4;q^4)_\infty^2(q^2;q^2)_\infty^2
\frac{(q^6r^2;q^4)_\infty
(1/q^2r^2;q^4)_\infty}{
(q^2/r;q^2)_\infty^2 (q^2r;q^2)_\infty^2}.
\end{eqnarray}
Let us now turn to the boundary expectation values of $\sigma_1^z$ in $H_B^{(\pm)}$. According to (4.42), it reduces to the known result for a diagonal boundary \cite{JKKKM}.  Namely,
\begin{eqnarray}
\frac{_B\langle 0;\pm|\sigma_1^z|\pm;0\rangle_B}{~_B\langle 0;\pm|\pm;0\rangle_B}
&=&
-1-2(1-r)^2\sum_{k=1}^\infty \frac{(-q^2)^k}{(1-rq^{2k})^2},
\\
\frac{_B\langle 1;\pm|\sigma_1^z|\pm;1\rangle_B}{~_B\langle 1;\pm|\pm;1\rangle_B}
&=&
-1-2(1-r)^2\sum_{k=1}^\infty \frac{(-q^2)^k}{(1-rq^{2k})^2}
\nonumber\\
&+&
2 (q^4;q^4)_\infty^2 (q^2;q^2)_\infty^2
\frac{(q^2r^2;q^4)_\infty 
(q^2/r^2;q^4)_\infty}{(q^2r;q^2)_\infty^2 (q^2/r;q^2)_\infty^2}.
\end{eqnarray}
Finally, note that the remaining expectation values are vanishing:
\begin{eqnarray}
\frac{~_B\langle i;\pm|
\sigma_1^\pm|\pm;i\rangle_B}
{~_B\langle i;\pm|\pm;i\rangle_B}
&=&0~~~~~{\rm for}~~i=0,1.
\end{eqnarray}
Simplifications occur upon the specializations $r=\pm 1, 0, \infty$: such cases correspond to  
free (Neumann) boundary condition $r=-1$ $(h=0)$,
fixed (Dirichlet) boundary condition $r=1$ $(h=\infty)$ whereas
for $r=0, \infty$ the Hamiltonian enjoys formal $U_q(sl_2)$ invariance.

\subsection{Relations between multiple integrals}

Linear relations between $n$-fold integrals are known in the mathematical literature \cite{TV,Ra,spi}. 
In the context of conformal field theory, some examples also arise in the calculation of correlation functions which contain screening operators. 
According to the spin-reversal property (\ref{eqn:reversal-correlation}), 
infinitely many relations of this kind can be exhibited based on 
previous results. Note that our relations can not be reduced to the relations between
$n$-fold integrals of elliptic gamma
functions summarized in \cite{Ra,spi}. 
Also, note that we understand the RHS of the spin-reversal property 
(\ref{eqn:reversal-correlation}) as an analytic continuation 
of the parameter $r$.
Here, we focus on the simplest examples:  a relation between a  triple integral and a single integral that has been computed explicitly in previous subsection is exhibited in two different cases. First, from
\begin{eqnarray}
P_{+,+}^{(+,0)}(-q^{-1}\zeta,\zeta;r,s)=P_{-,-}^{(-,1)}(-q^{-1}\zeta,\zeta;1/r,-s/r)
\end{eqnarray}
we have the following identity of multiple integrals of the elliptic theta function:
\begin{eqnarray}
&&r z q^2 
\frac{(q^2;q^2)_\infty^4}{(q^4;q^4)_\infty^2}
\frac{(1+r/z)(1-1/rz)(1-z/r)}{1-z^2}\frac{\Theta_{q^4}(z^2)\Theta_{q^4}(q^6r^2)}{
\Theta_{q^2}(q^2rz)\Theta_{q^2}(q^2r/z)}\nonumber\\
&+&
2+(1-z/r)
\sum_{k=1}^\infty (-1)^{k}
\frac{(z-z^{-1})q^{2k}/r+q^{2k}+q^{-2k}-(z+z^{-1})/r}{
(1-q^{2k}z/r)(1-q^{2k}/rz)}
\nonumber
\\
&=&
q^2 \frac{(q^2;q^2)_\infty^8}{(q^4;q^4)_\infty^4}
\frac{\Theta_{q^4}(z^2)}{1-z^2}
\left(
-q^2\int \int \int_{C_0^{(+,0)}}
+\int \int \int_{C_1^{(+,0)}}\right)
\prod_{a=1}^3 \frac{dw_a}{2\pi\sqrt{-1}}\frac{w_1}{w_2 w_3^5}\nonumber\\
&\times&
\frac{\displaystyle
(1-rz)(1-rw_3/q^3)\prod_{a=1}^2 (1-q^2/zw_a)}{
\displaystyle
\prod_{a=1}^2 (1-rw_a/q^2)(1-q^2w_1/w_2)(1-q^4/w_1w_2)}\nonumber\\
&\times&
\frac{\displaystyle
\Theta_{q^2}(w_1w_2)\Theta_{q^2}(w_2/w_1)
\Theta_{q^2}(zw_3/q)\Theta_{q^2}(qw_3/z)\prod_{a=1}^3
\Theta_{q^4}(w_a^2/q^2)}{
\displaystyle
\prod_{a=1}^2 \Theta_{q^2}(w_az)\Theta_{q^2}(w_a/z)
\Theta_{q^2}(w_3w_a/q)\Theta_{q^2}(w_a/qw_3)}.\label{eqn:theta1}
\end{eqnarray}
Here we have set $w_3=v$.
The integration contour $C_0^{(+,0)}$ is a simple closed curve such that
the $w_1$ encircles
$q^{2+2s}z,~q^{4+2s}/z,~q^{-1+2s}w_3,~q^{5+2s}/w_3$,
the $w_2$ encircles $q^{4+2s}z, q^{4+2s}/z, q^{-1+2s}w_3, q^{5+2s}/w_3$,
and the $w_3$ encircles $q^{3+2s}w_a, q^{5+2s}/w_a~(a=1,2)$,
for $s=0,1,2,\cdots$.
The integration contour $C_1^{(+,1)}$ is a simple closed curve such that
the $w_1$ encircles $q^{2+2s}z, q^{4+2s}/z, q^{3+2s}w_3, q^{5+2s}/w_3$,
the $w_2$ encircles $q^{4+2s}z, q^{4+2s}/z, q^{3+2s}w_3, q^{5+2s}/w_3$,
and $w_3$ encircles $q^{-1+2s}w_a, q^{5+2s}/w_a$ $(a=1,2)$,
for $s=0,1,2,\cdots$.

Secondly, from
\begin{eqnarray}
P_{+,+}^{(+,1)}(-q^{-1}\zeta,\zeta;r,s)=P_{-,-}^{(-,0)}(-q^{-1}\zeta,\zeta;1/r,-s/r)
\end{eqnarray}
we have the following identity:
\begin{eqnarray}
&&
2+\frac{1-z/r}{z}
\sum_{k=1}^\infty
(-q^2)^{k}\frac{
(z-z^{-1})-(1+q^{4k})/r+(z+z^{-1})q^{2k}
}{(1-q^{2k}z/r)(1-q^{2k}/rz)}
\nonumber\\
&=&q^2
\frac{(q^2;q^2)_\infty^8}{(q^4;q^4)_\infty^4}
\frac{\Theta_{q^4}(z^2)}{1-z^2}
\left(q^2 \int \int \int_{C_0^{(+,1)}}
-\int \int \int_{C_1^{(+,1)}}\right)
\prod_{a=1}^3 \frac{dw_a}{2\pi \sqrt{-1}}
\nonumber\\
&\times&
\frac{
\displaystyle
(1-1/rz)(1-q/rw_3)\prod_{a=1}^2(1-q^2/zw_a)}{
\displaystyle 
w_2^2 w_3^3 (1-q^2w_1/w_2)(1-q^4/w_1w_2)
\prod_{a=1,2}(1-q^2/rw_a)
}\nonumber\\
&\times&
\frac{
\Theta_{q^2}(w_1w_2)
\Theta_{q^2}(w_2/w_1)
\Theta_{q^2}(zw_3/q)
\Theta_{q^2}(qw_3/z)
\displaystyle
\prod_{a=1}^3
\Theta_{q^4}(w_a^2/q^2)}{
\displaystyle
\prod_{a=1}^2
\Theta_{q^2}(w_aw_3/q^2)
\Theta_{q^2}(w_a/qw_3)
\Theta_{q^2}(w_az)\Theta_{q^2}(w_a/z)}.\label{eqn:theta2}
\end{eqnarray}
The integration contour $C_l^{(+,1)}$ is a simple closed curve such that
$w_a$ $(a=1,2)$ encircles $q^2/r$ in addition the same points as the integration contour $C^{(+,0)}_l$ does.
The integration contour $C^{(+,0)}_l$ is given below (\ref{eqn:theta1}).
Obviously, using the spin-reversal property of the correlation functions,
we can write down infinitely many identities between multiple integrals of elliptic theta functions.

\section{Conclusion}

In the present paper, based on the $q$-vertex operator approach developed in
\cite{DFJMN, JKKKM}, two integral representations for 
the correlation functions of the half-infinite $XXZ$ spin chain with a triangular boundary have been derived 
in the massive regime.
In the special case of diagonal boundary condition,
known results are recovered.
Due to the presence of a non-diagonal boundary field coupled to the system,
the number of particle excitations is no longer conserved in this model. Here,  expectation values of the spin operators $\sigma_1^\pm$
which characterize this phenomena have been explicitly proposed.
Note that accordingly to \cite{JKM}, the analysis presented here may be extended to the massless regime
in a similar way. For the diagonal case, integral representations of the correlation functions are already known \cite{Kojima3,KZMNST}.
As mentioned in the Introduction, for a triangular boundary the construction of the transfer matrix' eigenvectors within the BA approach has been recently achieved \cite{PL}. Based on it, an alternative derivation of the correlation functions here presented would be highly desirable.

Although our results provide the first examples of correlation functions computed in the case of {\it non-diagonal} boundary conditions, one of the most interesting open problem is to extend the present analysis to the case of general integrable non-diagonal boundary condition. Indeed, for a finite size and generic boundary conditions,
due to the absence of reference states the algebraic Bethe ansatz cannot be applied.
In this case, alternative approaches are necessary.
At least, in the thermodynamic limit the above mentioned arguments about 
the structure of the eigenvectors in relation with the $q$-Onsager
algebra representation theory hold.
In this case, the explicit construction of monomials in terms of the $q$-Onsager generators can be also considered,
opening the way for the computation of correlation functions within
the $q$-vertex operator approach. Also, we would like to draw the reader's attention to other interesting open problems. It would be interesting to extend the analysis to models with higher symmetry - for instance $U_q(\widehat{sl}(M|N))$ - or the ABF model governed by the elliptic quantum group  ${\cal B}_{q,\lambda}(\widehat{sl_2})$ \cite{Baxter2,MW}.  Besides, having a better understanding of the space of states through the representation theory is highly desirable. Indeed, extracting interesting physical data - except in some special cases - from integral representations of correlation functions is a rather complicated problem. In this direction, the remarkable connection between the $q$-Onsager algebra and the theory of special functions \cite{Ter03} may be promising, as well as the link between solutions of the reflection quantum Knizhnik--Zamolochikov equations \cite{JKKKM} 
and  Koornwinder polynomials \cite{Ka,Stok} (see also \cite{DFZ}).

Finally, we would like to point out that promising routes have been explored recently. Within
Sklyanin's framework, let us mention for instance the functional approach of Galleas \cite{Galleas}, the extension of Sklyanin's separation of variable approach \cite{Niccoli} or the modified algebraic Bethe ansatz approach proposed in
\cite{BC} which may provide an alternative derivation of above results.

\section*{Acknowledgements}

The authors would like to thank  S.Belliard, V.Fateev, K.Kozlowski,
J.M.Maillet, G.Niccoli, and Y-Z.Zhang.
P.B. thanks J.Stokman for pointing out reference \cite{Ra}, and P.Zinn-Justin 
for discussions. T.K. 
  thanks S.Tsujimoto for discussions. 
  T.K. would like to thank Laboratoire de Math\'ematiques et Physique Th\'eorique, Universit\'e de Tours
for kind invitation and warm hospitality during his stay in March 2013.
This work is supported by the Grant-in-Aid for Scientific Research {\bf C} (21540228) from JSPS and Visiting professorship from CNRS.

\begin{appendix}

\section{Quantum group $U_q(\widehat{sl_2})$}

\label{appendix:A}

In this Appendix we recall the definition of $U_q(\widehat{sl_2})$
\cite{Jimbo, Drinfeld} and fix the notation that are used in the main text.
Let $-1<q<0$. Consider a free Abelian group on the letters 
$\Lambda_0, \Lambda_1, \delta$.
We call $P={\bf Z}\Lambda_0 \oplus {\bf Z}\Lambda_1 \oplus {\bf Z}\delta$ the weight lattice,
$\Lambda_i$ the fundamental weights and $\delta$ the null root. 
Define the simple roots $\alpha_i$ $(i=0,1)$ and the element $\rho$ by
$\alpha_0+\alpha_1=\delta$, $\Lambda_1=\Lambda_0+\frac{\alpha_1}{2}$, $\rho=\Lambda_0+\Lambda_1$.
Let $(h_0,h_1,d)$
be an ordered basis of $P^*=Hom(P,{\bf Z})$ dual to $(\Lambda_0,\Lambda_1,\delta)$.
We define a symmetric bilinear form $(,) : P\times P \to \frac{1}{2}{\bf Z}$ by
\begin{eqnarray}
(\Lambda_0, \Lambda_0)=0,~(\Lambda_0,\alpha_1)=0,~(\Lambda_0, \delta)=1,~ (\alpha_1,\alpha_1)=2,~
(\alpha_1,\delta)=0,~ (\delta,\delta)=0.
\end{eqnarray}
Regarding $P^* \subset P$ via this bilinear form we have the identification
\begin{eqnarray}
h_0=\alpha_0,~~h_1=\alpha_1,~~d=\Lambda_0.
\end{eqnarray}
The quantum group $U_q(\widehat{sl_2})$ is a $q$-analogue of the universal enveloping algebra
$U(\widehat{sl_2})$ generated by Chevalley generators $e_j, f_j$ $(j=0,1)$ and $q^h$ $(h \in P^*)$
and through the defining relations:
\begin{eqnarray}
&&~q^0=1,~~q^{h}q^{h'}=q^{h+h'},\\
&&~q^h e_j q^{-h}=q^{(h,\alpha_j)}e_j,~~q^h f_j q^{-h}=q^{-(h,\alpha_j)}f_j,\\
&&~[e_i,f_j]=\delta_{i,j}\frac{q^{h_i}-q^{-h_i}}{q-q^{-1}},\\
&&~e_i^3 e_j-[3]_q e_i^2 e_j e_i+[3]_q e_i e_j e_i^2-e_j e_i^3=0~~~(i \neq j),\\
&&~f_i^3 f_j-[3]_q f_i^2 f_j f_i+[3]_q f_i f_j f_i^2-f_j f_i^3=0~~~(i \neq j).
\end{eqnarray}
Here we denote:
\begin{eqnarray}
~[n]_q=\frac{q^n-q^{-n}}{q-q^{-1}}.
\end{eqnarray}
The quantum group $U_q(\widehat{sl_2})$ has the coproduct $\Delta$ structure:
\begin{eqnarray}
\Delta(q^h)=q^h \otimes q^h,~~\Delta(e_j)=e_j \otimes 1+q^{h_j}\otimes e_j,~~
\Delta(f_j)=f_j\otimes q^{-h_j}+1 \otimes f_j.
\end{eqnarray}
The coproduct $\Delta$ satisfies an algebra automorphism $\Delta(XY)=\Delta(X)\Delta(Y)$.

The quantum group $U_q(\widehat{sl_2})$ has another realization called the Drinfeld's second realization.
The generators of the Drinfeld's realizations are 
\begin{eqnarray}
a_m~(m \in {\bf Z}_{\neq 0}),~~x_m^\pm~(m \in {\bf Z}),~\gamma^{\frac{1}{2}},~~K,~~q^d.
\end{eqnarray}
In order to write down the defining relations, 
it is convenient to introduce the generating function
\begin{eqnarray}
X^\pm(z)&=&\sum_{m \in {\bf Z}}x_m^\pm z^{-m-1},\\
\psi^+(z)&=&K\exp\left((q-q^{-1})\sum_{n=1}^\infty a_n z^{-n}\right),\\
\psi^-(z)&=&K^{-1}\exp\left(-(q-q^{-1})\sum_{n=1}^\infty a_{-n}z^n\right).
\end{eqnarray}
Defining relations are given by
\begin{eqnarray}
&&[\gamma^{\frac{1}{2}}, U_q(\widehat{sl_2})]=0,\\
&&[a_m,a_n]=\delta_{m+n,0}\frac{[2m]_q}{m}\frac{\gamma^m-\gamma^{-m}}{q-q^{-1}},
\label{def:Drinfeld-boson}
\\
&&K a_m K^{-1}=a_m,~~~K X^\pm(z)K^{-1}=q^{\pm 2}X^\pm(z),\\
&&[a_m,X^\pm(z)]=\pm \frac{[2m]_q}{m}\gamma^{\mp |m|/2}z^m X^\pm(z),\\
&&(z-q^{\pm 2}w)X^\pm(z)X^\pm(w)=(q^{\pm 2}z-w)X^\pm(w)X^\pm(z),
\end{eqnarray}
\begin{eqnarray}
~[X^+(z),X^-(w)]&=&\frac{1}{(q-q^{-1})zw}
\left( 
\psi^+(\gamma^{\frac{1}{2}}w)\delta(z/\gamma w)
-
\psi^-(\gamma^{-\frac{1}{2}}w)\delta(\gamma z/w)\right).
\end{eqnarray}
Here $\delta(z)=\sum_{n \in {\bf Z}}z^n$ is a formal power series.
Other defining relations are given by
\begin{eqnarray}
q^d \gamma^{\frac{1}{2}} q^{-d}=\gamma^{\frac{1}{2}},~~
q^d K q^{-d}=K,~~
q^d x_m^\pm q^{-d}=q^m x_m^\pm,~~q^d a_m q^{-d}=q^m a_m.
\end{eqnarray}
The Chevalley generators are related to the Drinfeld generators as follows:
\begin{eqnarray}
K=q^{h_1},~~x_0^+=e_1,~~x_0^-=f_1,~~
\gamma K^{-1}=q^{h_0},~~x_1^-=e_0 q^{h_1},~~x_{-1}^+=q^{-h_1}f_0.
\end{eqnarray}
The Drinfeld realization is convenient to study bosonizations.

\section{Bosonization}

\label{appendix:B}

In this Appendix we review
the bosonizations of the quantum group $U_q(\widehat{sl_2})$ and the $q$-vertex operators for level $k=1$ \cite{DFJMN, FJ}.
The center $\gamma^{\frac{1}{2}}$ of $U_q(\widehat{sl_2})$ satisfies $(\gamma^{\frac{1}{2}})^2=q^{h_0+h_1}=q^k$
on the level $k$ representation.
Hence $\gamma=q$ on the level $k=1$ representation.
For the level $k=1$ case,
the defining relation of $a_m~(m \in {\bf Z}_{\neq 0})$ (\ref{def:Drinfeld-boson}) becomes
\begin{eqnarray}
[a_m,a_n]=\delta_{m+n,0}\frac{[2m]_q[m]_q}{m}~~~(m,n\neq 0).
\end{eqnarray}
We introduce the zero-mode operator $\partial, \alpha$ by $[\partial,\alpha]=2$ and
the normal ordering symbol $:~:$
\begin{eqnarray}
:a_m a_n :=\left\{\begin{array}{cc}
a_m a_n &~(m<0)\\
a_n a_m &~(m>0)
\end{array}\right.,
~~~:\alpha \partial:=:\partial \alpha:=\alpha \partial.\label{def:normal-ordering}
\end{eqnarray}
The bosonization of the irreducible highest representation $V(\Lambda_i)$
with fundamental weights $\Lambda_i$
are given by (\ref{eqn:bosonization-module}).
On this space the actions of $a_m$, $e^\alpha$, $\partial$
are given by
\begin{eqnarray}
&& 
a_m (f \otimes e^\beta)=\left\{
\begin{array}{cc}
~a_m f \otimes e^\beta &~(m<0),\\
~[a_m,f] \otimes e^\beta &~(m>0),
\end{array}
\right.
\label{def:bosonization-action-1}\\
&&
e^\alpha (f \otimes e^\beta)=f \otimes e^{\alpha+\beta},~~
\partial (f \otimes e^\beta)=(\alpha, \beta)f \otimes e^\beta,
\label{def:bosonization-action-2}
\end{eqnarray}
where $f \in {\bf C}[a_{-1}, a_{-2}, \cdots]$, and
we have set $[\partial, \Lambda_0]=0$ and $\Lambda_1=\Lambda_0+\frac{\alpha}{2}$.

The action of other Drinfeld generators is given by
\begin{eqnarray}
&&K=q^{\partial},~~\gamma=q,\\
&&X^\pm(z)=\exp\left(R^\pm(z)\right)\exp\left(S^\pm(z)\right)e^{\pm \alpha}z^{\pm \partial},\\
&&q^d (1 \otimes e^{\Lambda_i+n \alpha})=q^{-(\beta,\beta)/2+i/4}(1\otimes e^{\Lambda_i+n \alpha}),
\end{eqnarray}
where 
\begin{eqnarray}
R^\pm(z)=\pm \sum_{n=1}^\infty \frac{a_{-n}}{[n]_q}q^{\mp n/2}z^n,~~~
S^\pm(z)=\mp\sum_{n=1}^\infty \frac{a_n}{[n]_q}q^{\mp n/2}z^{-n}.
\end{eqnarray}
We have the following bosonizations of the $q$-vertex operators.
\begin{eqnarray}
\Phi_-^{(1-i,i)}(\zeta)&=&\exp\left(P(\zeta^2)\right)\exp\left(Q(\zeta^2)\right)e^{\alpha/2}
(-q^3\zeta^2)^{(\partial+i)/2}\zeta^{-i},
\label{boson:VO1}
\\
\Phi_+^{(1-i,i)}(\zeta)&=&
\oint_{\widetilde{C}_1} \frac{dw}{2\pi\sqrt{-1}}\frac{(1-q^2)w\zeta}{
q(w-q^2\zeta^2)(w-q^4\zeta^2)}:\Phi_-^{(1-i,i)}(\zeta)X^-(w):,
\label{boson:VO2},
\end{eqnarray}
\begin{eqnarray}
\Psi_-^{* (1-i,i)}(\xi)&=&\exp\left(-P(q^{-1}\xi^2)\right)
\exp\left(-Q(q\xi^2)\right)e^{-\alpha/2}(-q^3\xi^2)^{(-\partial+i)/2}\xi^{1-i},\\
\Psi_+^{* (1-i,i)}(\xi)&=&
\oint_{\widetilde{C}_2}\frac{dw}{2\pi\sqrt{-1}}
\frac{q^2(1-q^2)\xi}{(w-q^2\xi^2)(w-q^4\xi^2)}:\Psi_-^{* (1-i,i)}(\xi)X^+(w):,
\end{eqnarray}
where
\begin{eqnarray}
P(z)=\sum_{n=1}^\infty
\frac{a_{-n}}{[2n]_q}q^{7n/2}z^n,~~~ Q(z)=-\sum_{n=1}^\infty \frac{a_n}{[2n]_q}q^{-5n/2}z^{-n}.
\end{eqnarray}
Here the integration contour $\widetilde{C}_1$ is a simple closed curve such that
the $w$ encircles $q^{4}\zeta^2$ inside but not $q^2\zeta^2$.
The integration contour $\widetilde{C}_2$ is a simple closed curve such that
the $w$ encircles $q^{2}\zeta^2$ inside but not $q^4 \zeta^2$.

In what follows we summarize normal orderings.
\begin{eqnarray}
&&
\Phi_-^{(i,1-i)}(\zeta_1)\Phi_-^{(1-i,i)}(\zeta_2)=
(-q^3 \zeta_1^2)^{\frac{1}{2}}
\frac{(q^2 \zeta_2^2/\zeta_1^2 ; q^4)_\infty}{
(q^4 \zeta_2^2/\zeta_1^2; q^4)_\infty}
:\Phi_-^{(i,1-i)}(\zeta_1)\Phi_-^{(1-i,i)}(\zeta_2):,
\\
&&
\Psi_-^{* (i,1-i)}(\xi_1)\Psi_-^{* (1-i,i)}(\xi_2)=
(-q^3 \xi_1^2)^{\frac{1}{2}}
\frac{(\xi_2^2/\xi_1^2;q^4)_\infty}{
(q^2 \xi_2^2/\xi_1^2;q^4)_\infty}:
\Psi_-^{* (1-i,i)}(\xi_1)\Psi_-^{* (1-i,i)}(\xi_2)
:,
\\
&&
\Phi_-^{(i,1-i)}(\zeta)\Psi_-^{* (1-i,i)}(\xi)=
(-q^3 \zeta^2)^{-\frac{1}{2}}
\frac{(q^3\xi^2/\zeta^2;q^4)_\infty}{
(q\xi^2/\zeta^2;q^4)_\infty}
:\Phi_-^{(i,1-i)}(\zeta)\Psi_-^{* (1-i,i)}(\xi):,
\\
&&\Psi_-^{* (i,1-i)}(\xi)\Phi_-^{(1-i,i)}(\zeta)=
(-q^3 \xi^2)^{-\frac{1}{2}}\frac{
(q^3 \zeta^2/\xi^2;q^4)_\infty}{(q \zeta^2/\xi^2;q^4)_\infty}
:
\Psi_-^{* (i,1-i)}(\xi)\Phi_-^{(1-i,i)}(\zeta):,
\\
&&X^+(w_1)X^+(w_2)=w_1^2 (1-w_2/w_1)(1-w_2/q^2 w_1):
X^+(w_1)X^+(w_2):,
\\
&&X^-(w_1)X^-(w_2)=w_1^2 (1-w_2/w_1)(1-q^2 w_2/w_1):
X^-(w_1)X^-(w_2)
:,
\\
&&X^+(w_1)X^-(w_2)=\frac{1}{w_1^2 (1-q w_2/w_1)(1-w_2/q w_1)}
:
X^+(w_1)X^-(w_2)
:,
\\
&&X^-(w_1)X^+(w_2)=\frac{1}{w_1^2 (1-q w_2/w_1)(1-w_2/q w_1)}
:
X^-(w_1)X^+(w_2)
:,
\\
&&\Phi_-^{(i,1-i)}(\zeta)X^+(w)=(-q^3 \zeta^2)(1-w/q^3\zeta^2)
:
\Phi_-^{(i,1-i)}(\zeta)X^+(w)
:,
\\
&&X^+(w)\Phi_-^{(i,1-i)}(\zeta)=w(1-q^3 \zeta^2/w)
:
X^+(w)\Phi_-^{(i,1-i)}(\zeta)
:,
\\
&&\Phi_-^{(i,1-i)}(\zeta)X^-(w)=\frac{-1}{q^3 \zeta^2 (1-w/q^2 \zeta^2)}
:
\Phi_-^{(i,1-i)}(\zeta)X^-(w)
:,
\\
&&X^-(w)\Phi_-^{(i,1-i)}(\zeta)=
\frac{1}{w(1-q^4 \zeta^2/w)}
:
X^-(w)\Phi_-^{(i,1-i)}(\zeta)
:,
\\
&&
\Psi_-^{* (i,1-i)}(\xi)X^+(w)
=\frac{-1}{q^3 \xi^2 (1-w/q^4 \xi^2)}
:
\Psi_-^{* (i,1-i)}(\xi)X^+(w)
:,
\\
&&
X^+(w)\Psi_-^{* (i,1-i)}(\xi)=\frac{1}{w (1-q^2 \xi^2/w)}
:
X^+(w)\Psi_-^{* (i,1-i)}(\xi)
:,
\\
&&
\Psi_-^{* (i,1-i)}(\xi)X^-(w)=(-q^3 \xi^2)
(1-w/q^3 \xi^2)
:
\Psi_-^{* (i,1-i)}(\xi)X^-(w)
:,
\\
&&
X^-(w)\Psi_-^{* (i,1-i)}(\xi)=w (1-q^3 \xi^2/w)
:
X^-(w)\Psi_-^{* (i,1-i)}(\xi):.
\end{eqnarray}

\section{Convenient Formulae}

\label{appendix:C}

In this Appendix we summarize convenient formulae for calculations of
vacuum expectation values.
\begin{eqnarray}
\exp\left(\sum_{n=1}^\infty
\frac{1}{(q-q^{-1})}\frac{q^{(c+2)n}}{n[2n]_q}y^n
\right)&=&
(q^{c+4}y;q^4)_\infty,\\
\exp\left(\sum_{n=1}^\infty
\frac{1}{(q-q^{-1})}\frac{[n]_q q^{(c+2)n}}{n[2n]_q^2}y^n
\right)&=&\frac{
(q^{c+5}y;q^4,q^4)_\infty}{(q^{c+7}y;q^4,q^4)_\infty},\\
\exp\left(\sum_{n=1}^\infty
\frac{1}{(q-q^{-1})}\frac{q^{(c+2)n}}{n[n]_q}y^n
\right)&=&
(q^{c+3}y;q^4)_\infty,\\
\exp\left(\sum_{n>0}\frac{[n]_q}{n[2n]_q}\frac{q^{cn}}{1-\alpha_n \gamma_n}
y^n\right)
&=&
\frac{(q^{c+3}y;q^4,q^4)_\infty}{
(q^{c+1}y;q^4,q^4)_\infty},\\
\exp\left(
\sum_{n>0}\frac{[2n]_q}{n[n]_q}\frac{q^{cn}}{1-\alpha_n \gamma_n}y^n
\right)&=&\frac{1}{(q^{c-1}y;q^2)_\infty}.
\end{eqnarray}
\begin{eqnarray}
\exp\left(\sum_{n=1}^\infty \frac{[n]_q}
{n(1-\alpha_n\gamma_n)}\delta_n^{(i)}q^{cn/2}y^n\right)&=&\left(
\frac{(q^{c+3}y^2;q^8,q^8)_\infty
(q^{c+1}y^2;q^8,q^8)_\infty}{
(q^{c-1}y^2;q^8,q^8)_\infty
(q^{c+5}y^2;q^8,q^8)_\infty}
\right)^{1/2}\nonumber\\
&\times&
\left(\frac{(q^{(c-3)/2+2i}r^{1-2i}y;q^4,q^4)_\infty}{
(q^{(c+1)/2+2i}r^{1-2i}y;q^4,q^4)_\infty}\right)^{1-2i},\\
\exp\left(\sum_{n=1}^\infty \frac{[2n]_q}
{n(1-\alpha_n\gamma_n)}\delta_n^{(i)}q^{cn/2}y^n\right)&=&
\left(\frac{(q^{c-1}y^2;q^8)_\infty}{(q^{c-3}y^2;q^8)_\infty}\right)^{1/2}
(q^{(c-5)/2+2i} r^{1-2i}y;q^4)_\infty^{1-2i},\\
\exp\left(\sum_{n=1}^\infty \frac{[n]_q}
{n(1-\alpha_n\gamma_n)}
\beta_n^{(i)}q^{cn/2}y^n\right)&=&\left(
\frac{(q^{c+7}y^2;q^8,q^8)_\infty
(q^{c+13}y^2;q^8,q^8)_\infty}{
(q^{c+9}y^2;q^8,q^8)_\infty
(q^{c+11}y^2;q^8,q^8)_\infty}
\right)^{1/2}\nonumber\\
&\times&
\left(\frac{(q^{(c+9)/2-2i}r^{1-2i}y;q^4,q^4)_\infty}{
(q^{(c+13)/2-2i}r^{1-2i}y;q^4,q^4)_\infty}\right)^{1-2i},\\
\exp\left(\sum_{n=1}^\infty \frac{[2n]_q}
{n(1-\alpha_n\gamma_n)}\beta_n^{(i)}q^{cn/2}y^n\right)&=&
\left(\frac{(q^{c+5}y^2;q^8)_\infty}{(q^{c+7}y^2;q^8)_\infty}\right)^{1/2}
(q^{(c+7)/2-2i} r^{1-2i}y;q^4)_\infty^{1-2i}.
\end{eqnarray}

~\\

\end{appendix}

\end{document}